\documentclass[aps,pra,twocolumn,showpacs,10pt]{revtex4-1}
\usepackage{amsmath,amsthm,amssymb,listings,color,pgfplotstable,pgfplots,tikz}
\pgfplotsset{compat=1.14}
\usetikzlibrary{positioning,patterns}

\tikzset{
  pics/box/.style args={#1/#2/#3/#4/#5/#6/#7/#8}{
    code = {
      \coordinate[name prefix .., below left=#2 and -#3 of circuit.north west] (#1coord);
      \node[rounded corners, fill=black, opacity=0.2, minimum height=#4,minimum width=#5] at (coord) (theBox) {};
      \node[below right=#7 and #8 of theBox.north west, font=\scriptsize\itshape] (text) {#6};
    }
  },
  pics/opaquebox/.style args={#1/#2/#3/#4/#5/#6}{
    code = {
      \coordinate[name prefix .., below left=#2 and -#3 of circuit.north west] (#1coord);
      \node[rounded corners, font=\small\itshape, fill=black!20, opacity=1, minimum height=#4,minimum width=#5] at (coord) (theBox) {#6};
    }
  },
  pics/cut/.style args={#1/#2}{
    code = {
      \draw[name prefix .., dash pattern={on 7pt off 2pt on 1pt off 2pt}, line width=1pt, opacity=0.65] (#1) -- (#2);
    }
  },
  pics/ebit/.style args={#1/#2/#3}{
    code = {
      \node[name prefix .., below left=#2 and -#3 of circuit.north west, rectangle, fill=white, minimum height=10mm, minimum width=#3] (#1clear) {};
      \coordinate[below left=1.5mm and 0mm of clear.north east] (one);
      \coordinate[below=7.04mm of one] (two);
      \draw[name prefix ..] (#1one) edge[out=180,in=180,looseness=1.5] (#1two);
    }
  },
  pics/ebitLessClear/.style args={#1/#2/#3}{
    code = {
      \node[name prefix .., below left=#2 and -#3 of circuit.north west, rectangle, fill=white, minimum height=10mm, minimum width=5mm] (#1clear) {};
      \coordinate[below left=1.5mm and 0mm of clear.north east] (one);
      \coordinate[below=7.04mm of one] (two);
      \draw[name prefix ..] (#1one) edge[out=180,in=180,looseness=1.5] (#1two);
    }
  },
  pics/ebitLong/.style args={#1/#2/#3}{
    code = {
      \node[name prefix .., below left=#2 and -#3 of circuit.north west, rectangle, fill=white, minimum height=25mm, minimum width=#3] (#1clear) {};
      \coordinate[below left=1.5mm and 0mm of clear.north east] (one);
      \coordinate[below=21.12mm of one] (two);
      \draw[name prefix ..] (#1one) edge[out=180,in=180,looseness=1.5] (#1two);
    }
  },
}

\lstnewenvironment{algorithm}{
    \lstset{ 
        mathescape=true,
        frame=tb,
        numbers=left, 
        numberstyle=\tiny,
        basicstyle=\scriptsize, 
        keywordstyle=\bfseries\em,
        keywords={,input, output, and, or, if, then, else, foreach, in, do, begin,end,}
        morecomment=[l]{//}, 
        commentstyle=\itshape,
        numbers=left,
        xleftmargin=.04\textwidth
    }
}{}

\newcommand{\ket}[1]{{\lvert{#1}\rangle}}

\begin{document}

\title{Automated distribution of quantum circuits via hypergraph partitioning}

\author{Pablo Andr{\'e}s-Mart{\'i}nez}
\email[]{p.andres-martinez@ed.ac.uk}
\thanks{Supported by the CDT in Pervasive Parallelism, funded by the EPSRC (grant EP/L01503X/1) and the School of Informatics (University of Edinburgh). Some of the work was done during a visit to Dalhousie University; the visit was partially funded by the host institution.}

\author{Chris Heunen}
\email[]{chris.heunen@ed.ac.uk}
\thanks{Supported by EPSRC Research Fellowship EP/R044759/1. Both authors thank Petros Wallden, Peter Selinger and Neil Julien Ross. Discussions with Peter Selinger led to the extension of the algorithm described in Appendix B and improvements on the implementation. Neil Julien Ross suggested looking into CCZ gate distribution.}
\affiliation{University of Edinburgh}

\date{\today}

\begin{abstract}
Quantum algorithms are usually described as monolithic circuits, becoming large at modest input size. Near-term quantum architectures can only manage a small number of qubits.
We develop an automated method to distribute quantum circuits over multiple agents, minimising quantum communication between them.
We reduce the problem to hypergraph partitioning and then solve it with state-of-the-art optimisers. This makes our approach useful in practice, unlike previous methods.
Our implementation is evaluated on five quantum circuits of practical relevance. 
\end{abstract}

\maketitle

\section{Introduction}

Quantum computation~\cite{feynman,nielsenchuang,lloyd:simulators} employs the laws of quantum mechanics to design systems capable of outperforming classical computers in certain problems~\cite{Shor,Grover,harrowhassadimlloyd:linear}. Over the past couple of decades, this idea has rapidly developed from theoretical results into actual quantum technology~\cite{obrienfurusawavuckovic:photonic,devoretschoelkopf:superconducting,horodecki:entanglement}.

\begin{figure}
    \begin{tikzpicture}
      \node[inner sep=0pt] (circuit) {\includegraphics[scale=1.8]{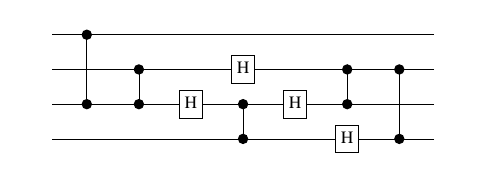}};  
      \node[above left=8mm and -9mm of circuit.west, opacity=0.9] {\footnotesize \(A\)};
      \node[above left=1.5mm and -9mm of circuit.west, opacity=0.9] {\footnotesize \(B\)};
      \node[below left=1.5mm and -9mm of circuit.west, opacity=0.9] {\footnotesize \(C\)};
      \node[below left=8mm and -9mm of circuit.west, opacity=0.9] {\footnotesize \(D\)};
      \node[above right=4mm and 15.5mm of circuit.west, opacity=0.9] {\footnotesize \(\alpha\)};
      \node[right=25.5mm of circuit.west, opacity=0.9] {\footnotesize \(\beta\)};
      \node[below right=4.5mm and 44mm of circuit.west, opacity=0.9] {\footnotesize \(\gamma\)};
      \node[right=63mm of circuit.west, opacity=0.9] {\footnotesize \(\delta\)};
      \node[right=73mm of circuit.west, opacity=0.9] {\footnotesize \(\eta\)};
    \end{tikzpicture}
  \caption{Example quantum circuit, applying Hadamard gates $H$ and CZ gates $\alpha,\beta,\gamma,\delta$ and $\eta$ to four input qubits $A,B,C,D$.}
  \label{fig:circuit}
\end{figure}
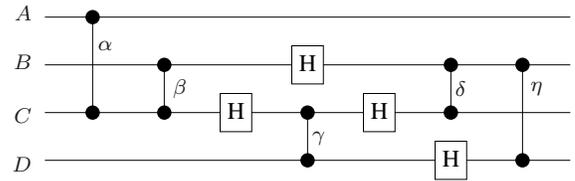

Although there are other approaches~\cite{raussendorfbriegel:mbqc}, the dominant way to present a quantum algorithm is as a quantum circuit~\cite{deutsch:networks}: a description of how quantum devices, chosen from a fixed finite set, are applied to different parts of the input system; see Figure~\ref{fig:circuit} for an example. Each of the `wires' that quantum devices act upon typically consist of a two level quantum system called a quantum bit or qubit.
The qubit count grows with the input size, and for relevant problems such as the \emph{unique shortest vector} problem (with applications in cryptography~\cite{USV}) the circuit grows large: lattice dimension 3 already requires 842 qubits and 95,624 gates~\footnote{Extracted from Quipper USV-R implementation~\cite{quipperonline}.}.

Near-term quantum computing architectures are not capable of executing such large circuits. We are currently entering the era of Noisy Intermediate-Scale Quantum (NISQ) technology~\cite{NISQ}, being able to fabricate small quantum computing units (QPU for short) ranging from \(10\) to almost \(100\) qubits. Much effort is being dedicated to further increase the number of qubits that QPUs can manage, but as the number of qubits grows, the challenge of addressing each qubit individually and shielding them from unwanted interactions and decoherence rapidly becomes unmanageable~\cite{vanmeterdevitt:survey}. To scale up beyond this point, researchers are proposing distributed quantum architectures~\cite{HierarchichalIonTrap,HierarchichalQuantumDot} that integrate multiple smaller QPUs that cooperate to simulate a larger circuit. This requires QPUs to coordinate, making it necessary to allocate resources for communication and establishing a trade-off: the more processors we wish to use to perform the computation, the larger the communication cost will be. In the extreme case, one could imagine each individual qubit being managed by a separate QPU, so every multi-qubit gate would require inter-QPU communication. 

Quantum communication is performed more profitably by photonics, whereas in-processor communication is easier with cold matter or solid state architectures. Let us mention two examples:
\begin{itemize}
  \item Some of the currently most advanced quantum architectures are hybrids, that connect small units of matter degrees of freedom (such as quantum dots, ion traps or nitrogen vacancy centres), into a network using photonic degrees of freedom~\cite{nqit,HierarchichalQuantumDot,delft}.
  \item Part of the aim of the Quantum Internet Alliance~\cite{quantuminternetalliance} is to establish a network between several parties, each of whose nodes have limited quantum capabilities in the order of 10-20 qubits~\cite{wehnerelkousshanson:quantuminternet}. In this view, questions of routing information along the quantum network become important~\cite{schouteetal:routing,hahnpappeeisert:routing}.
\end{itemize}

Although distributed quantum computing is being discussed for scalability purposes~\cite{vanmeterdevitt:survey}, and experimental distributed architectures have been proposed~\cite{HierarchichalIonTrap,HierarchichalQuantumDot}, the standard approach of quantum programmers remains designing quantum algorithms as monolithic circuits. But how do you execute, for example, the quantum circuit in Figure~\ref{fig:circuit}, using a pair of 2-qubit QPUs?
This document develops an automated method that distributes any circuit across any number of quantum processing units, while minimising the quantum communication between them. We reduce the problem to hypergraph partitioning, which has been extensively researched in computer science literature and has fast heuristic solvers~\cite{KaHyPar, PaToH}.

We first discuss distributed quantum computing in more detail, then we reduce the problem of distributing a circuit across multiple QPUs to hypergraph partitioning. 
We briefly discuss some implementation details, namely pre- and post-processing routines to improve the circuit distribution. Finally, we evaluate our results on five quantum circuits from the literature that are of interest to quantum computing.

\section{Distributed Quantum Computing}

This section describes the essential characteristics of any distributed architecture and identifies communication across QPUs as the main bottleneck. We then detail the problem at hand and discuss related work from the literature. Finally, we describe the standard way nonlocal multi-qubit gates are executed across QPUs, which will be required when building distributed circuits.

\subsection{Distributed quantum architectures} 
We claim that any distributed quantum architecture (DQA for short) shares the following essential features:
\begin{itemize}
  \item \emph{Multiple QPUs}, each of which holds a limited number of workspace qubits. It should be possible to prepare these qubits to hold the input data of a program and read output from them through measurements.

  \item A \emph{classical communication network} that the QPUs may send classical messages through when measuring their qubits and receive message over when applying corrections.

  \item \emph{Ebit generation hardware}. An \emph{ebit} is a maximally entangled bipartite quantum state shared across two QPUs. In this paper, we choose the Bell state \(\frac{1}{\sqrt{2}}\ket{00}+\frac{1}{\sqrt{2}}\ket{11}\) as the initial state for every ebit. Each ebit comprises two qubits, called \emph{ebit halves}, that are stored in different QPUs. An ebit should be understood as a resource that a QPU may use to communicate a single qubit to another QPU. 
  Each QPU may have its own hardware to create and share ebits, or a separate device may generate ebits centrally. Depending on the technology, this may involve entanglement distillation and error correction of a noisy quantum channel~\cite{bennettdivincenzosmolinwooters:distillation, NoisyChannels}. A promising way to create ebits is to excite the qubits we wish to entangle so that they each release a photon, which are then made to interfere at a beam splitter; the outcome of the interference heralds the creation of entanglement between the qubits~\cite{EbitGenHeralded}.
  
  \item \emph{Ebit memory space} on each QPU, dedicated to the storage (and possibly generation) of ebit halves. These are the qubits that will interact with the rest of the QPUs, and are thus likely to require a different physical realisation than the rest of the qubits used as workspace for the computation. QPUs should support the application of two-qubit gates between ebit and workspace qubits, so that the entanglement can be spread within the QPU.
\end{itemize}

A DQA using ion traps has been proposed~\cite{HierarchichalIonTrap} where each ion trap holding up to \(100\) qubits acts as a QPU. Ebit generation is achieved either by creating Bell states locally in one QPU and shuttling one of the qubits towards another QPU, or by the photon-heralded entanglement generation routine previously described~\cite{EbitGenHeralded}. The authors argue that, to reduce undesired crosstalk, the ions used for ebit generation will likely need to be of a different atomic species than those used for workspace qubits. Moreover, while they assume the cost of classical communication to be negligible, the authors estimate that the generation of ebits will be roughly \(300\) times more costly than the application of a two-qubit gate locally within a QPU. 

Other DQAs have been proposed using different technologies; for instance, via semiconductor nanophotonics~\cite{HierarchichalQuantumDot} where qubits are encoded by quantum dots that interact with their nearest-neighbours within cavities. Each of these nearest-neighbour groups of quantum dots corresponds to the workspace qubits of a QPU, while entanglement across QPUs is generated by laser pulses through strategically positioned waveguides that affect the quantum dots closer to it. For this DQA too, the authors claim that ebit generation is the bottleneck of the architecture. In general, this is to be expected for any distributed architecture: entanglement across distant parties is harder to achieve than interactions between neighbouring qubits. Thus, our objective when implementing circuits on a distributed architecture will be to minimise the amount of ebits required; this is the focus of the present paper.

There is no clear choice of the technology to be used to implement the essential DQA features listed above. For generality's sake we therefore choose not to make any further assumptions on the specification of DQAs, ignoring details that could vary across different technologies.

\subsection{Nonlocal quantum gates}

Quantum circuits are built from devices known as gates that apply operations to the qubits (see Figure~\ref{fig:circuit}). A universal gateset is a collection of gates that can be used to implement any circuit up to arbitrary accuracy. The Solovay-Kitaev theorem~\cite{SolovayKitaev} ensures that any circuit can be approximated (up to arbitrary precision $\varepsilon$) using only gates from any chosen universal gateset. This translation is efficient (poly-logarithmic with respect to $1/\varepsilon$) both in time and circuit depth. Therefore, without loss of generality, we assume that every circuit we are tasked to distribute is built up from one-qubit gates and CZ gates exclusively. This gateset is known to be universal, as it is locally equivalent to the Clifford+$T$ gateset. We choose CZ over the more usual CNOT gate because of its symmetry on inputs, which simplifies some details of our algorithm.

When distributing a circuit, one-qubit gates require no communication across QPUs and are therefore trivial from our point of view. In contrast, communication is required to implement any CZ gate acting on a pair of qubits that live in different QPUs, in which case we call the gate \emph{nonlocal}. To distribute quantum circuits we need to be able to implement nonlocal CZ gates.

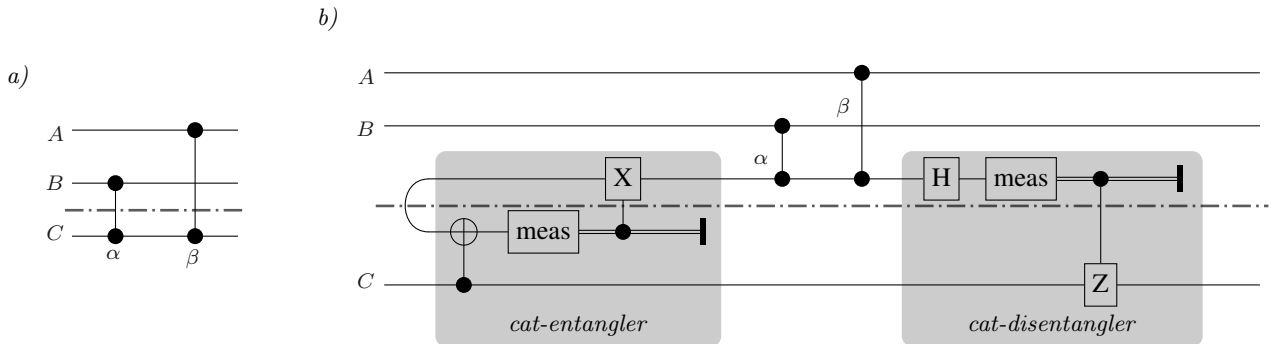
\begin{figure*}
\centering
\begin{tikzpicture}
  \node (distributed) {
    \begin{tikzpicture}
      \node[inner sep=0pt] (circuit) {\includegraphics[scale=2]{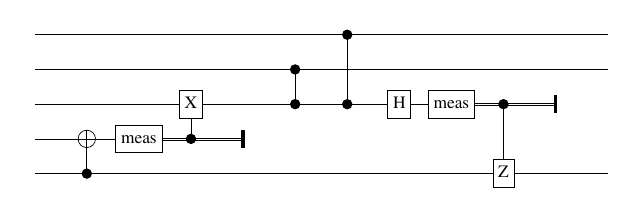}};
      \pic (e1) {ebit=e1/19.75mm/13mm};
      \coordinate[above left=-3.6mm and -6mm of circuit.west] (leftPoint);
      \coordinate[above right=-3.6mm and -6mm of circuit.east] (rightPoint);
      \pic (cut) {cut=leftPoint/rightPoint};
      \node[above left=11.5mm and -7mm of circuit.west, opacity=0.9] {\footnotesize \(A\)};
      \node[above left=4.5mm and -7mm of circuit.west, opacity=0.9] {\footnotesize \(B\)};
      \node[below left=11.5mm and -7mm of circuit.west, opacity=0.9] {\footnotesize \(C\)};
      \node[above right=0.8mm and 55.1mm of circuit.west, opacity=0.9] {\footnotesize \(\alpha\)};
      \node[above right=7.1mm and 66.1mm of circuit.west, opacity=0.9] {\footnotesize \(\beta\)};
      \node[rounded corners, minimum height=26mm, minimum width=38mm, fill=black, opacity=0.2, below left=-25mm and -52mm of circuit, label={[shift={(0,5.5mm)}]south:\textit{cat-entangler}}] (cat-entangler) {};
      \node[rounded corners, minimum height=26mm, minimum width=40mm, fill=black, opacity=0.2, below left=-25mm and -116mm of circuit, label={[shift={(0,5.5mm)}]south:\textit{cat-disentangler}}] (cat-disentangler) {};
      \node[right=-3mm of circuit.north west, font=\itshape] (text) {b)};
    \end{tikzpicture}
  };
  \node[left=0mm of distributed] (template) {
    \begin{tikzpicture}
      \node[inner sep=0pt] (circuit) {\includegraphics[trim={2.5mm 0 2.5mm 0}, clip=true, scale=2]{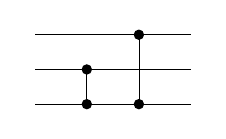}};  
      \node[left=-7mm of circuit.west, minimum height=20mm, minimum width=7mm, fill=white] (whiteLeft) {};
      \node[right=-7mm of circuit.east, minimum height=20mm, minimum width=7mm, fill=white] (whiteRight) {};
      \coordinate[below left=3.6mm and -6mm of circuit.west] (leftPoint);
      \coordinate[below right=3.6mm and -6mm of circuit.east] (rightPoint);
      \pic (cut) {cut=leftPoint/rightPoint};
      \node[above left=4.5mm and -7mm of circuit.west, opacity=0.9] {\footnotesize \(A\)};
      \node[left=-7mm of circuit.west, opacity=0.9] {\footnotesize \(B\)};
      \node[below left=4.5mm and -7mm of circuit.west, opacity=0.9] {\footnotesize \(C\)};
      \node[below right=7.7mm and 10.3mm of circuit.west, opacity=0.9] {\footnotesize \(\alpha\)};
      \node[below right=7.7mm and 20.9mm of circuit.west, opacity=0.9] {\footnotesize \(\beta\)};
      \node[right=-3mm of circuit.north west, font=\itshape] (text) {a)};
    \end{tikzpicture}
  };
\end{tikzpicture}
\caption{Implementation, as in~\cite{Nonlocal}, of a group of nonlocal CZ gates that share a common qubit. Circuit \textit{a)} is the original circuit, \textit{b)} is the distributed version. The dashed line indicates how the circuit is separated into two QPUs. The bent wire on the left of \textit{b)} represents an ebit. The measurement outcomes are communicated through a classical network.}
\label{fig:nonlocal}
\end{figure*}

We use an approach pioneered by Gottesman and Chuang~\cite{Gottesman} to implement multi-qubit gates across distant qubits using entangled mutipartite states and measurements. In particular, we will use the scheme proposed in~\cite{Nonlocal} which, using a single ebit, implements any number of contiguous nonlocal CZ gates that act on a common qubit, as shown in Figure~\ref{fig:nonlocal}. To do so, a so-called cat-entangler first shares the state of the common qubit with the remote QPU, which requires the use of an ebit; then, the CZ gates are locally performed on the remote QPU and finally the cat-disentangler destructively measures the remaining ebit half to remove any residual entanglement. 

Another option would be to use standard qubit teleportation~\cite{Teleportation} to send the qubit that the CZ gates share to the remote QPU; then the CZ gates may be applied locally, and afterwards the qubit can be sent back to its original QPU through another teleportation. The scheme in Figure~\ref{fig:nonlocal} uses a single ebit, whereas the teleportation approach just described would use two ebits to accomplish the same result, as each teleportation consumes an ebit. However, if it so happens that the teleported qubit is not required in its original QPU any more, we could skip the second teleportation and thus use a single ebit. This rather trivial remark is relevant in the discussion of Appendix B.

\subsection{The circuit distribution problem}

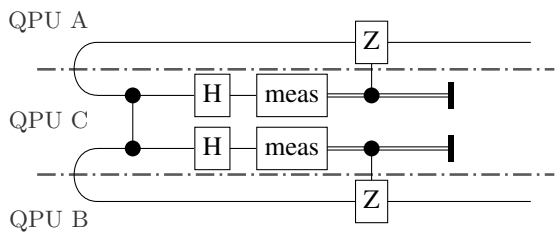
\begin{figure}
\centering
\begin{tikzpicture}
  \node[inner sep=0pt] (circuit) {\includegraphics[scale=2]{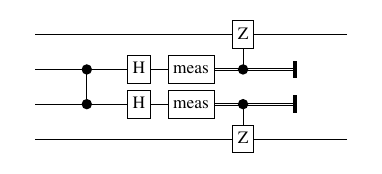}};
  \pic (e1) {ebit=e1/5.62mm/13mm};
  \pic (e2) {ebit=e2/19.75mm/13mm};
  \coordinate[above left=7mm and -5mm of circuit.west] (leftPoint1);
  \coordinate[above right=7mm and -4mm of circuit.east] (rightPoint1);
  \pic (cut1) {cut=leftPoint1/rightPoint1};
  \coordinate[below left=7mm and -5mm of circuit.west] (leftPoint2);
  \coordinate[below right=7mm and -4mm of circuit.east] (rightPoint2);
  \pic (cut2) {cut=leftPoint2/rightPoint2};
  \node[above left=-7mm and -13mm of circuit.north west, opacity=0.8] (A) {QPU A};
  \node[below left=-7mm and -13mm of circuit.south west, opacity=0.8] (B) {QPU B};
  \node[left=-13mm of circuit.west, opacity=0.8] (C) {QPU C};
\end{tikzpicture}
\caption{Initially, a pair of ebits (bent wires) entangle QPU A with C and QPU B with C. At the end of the circuit, the two unmeasured qubits form an ebit between QPU A and B. Two ebits and two classical communications are used.}
\label{fig:intermediary}
\end{figure}

Our objective is to \emph{minimise the number of ebits} required to distribute a given quantum circuit. 
We will assume ebits may be generated to entangle any pair of QPUs in the architecture, i.e.\ we consider no restriction on the ebit connectivity between QPUs. In practice, constraints are likely to exist but, as shown in Figure~\ref{fig:intermediary}, whenever two QPUs may not be directly entangled through the generation of an ebit, another QPU may act as intermediary and create the desired entanglement using two physically realisable ebits. A simple yet reasonable topological arrangement of QPUs is a hypercube, where \(N\) QPUs may be connected directly with \(log\ N\) other QPUs, and indirectly through at most \(log\ N\) physically realisable ebits by repeatedly using the method from Figure~\ref{fig:intermediary}. Although relevant, this \(log\ N\) factor (e.g.\ \(7\) ebits for a \(128\) QPU computer) is not the main bottleneck of the architecture; considering direct generation of ebits may already be up to \(300\) times more costly than local two-qubit operations (as estimated by the authors of the ion trap DQA~\cite{HierarchichalIonTrap}), the immediate concern is to reduce the overall use of ebits any distributed circuit requires, independently of whether each ebit can be generated directly or not.

On a similar note, we assume that the QPUs have no internal topological constraints, i.e.\ each one of them is capable of applying CZ gates upon any pair of qubits it holds; this idealisation can be accounted for at a later stage. Recent research has provided automated methods for efficiently simulating any circuit on topologically constrained QPUs, either by finding the least amount of qubit swappings required~\cite{ChildsSWAPs} or by redesigning the circuit from scratch using, for instance, Steiner trees~\cite{KissingerSteinerTrees,NashSteinerTrees}. In practice, any of these methods may be used to simulate each of the circuits our algorithm (see Section~\ref{sec:algorithm}) allocates to each QPU, so the QPUs may actually be topologically constrained.

Although similar at first glance, the problem we focus on (quantum circuit distribution, QCD for short) is fundamentally different from that of simulation of circuits on topologically constrained QPUs~\cite{ChildsSWAPs,KissingerSteinerTrees,NashSteinerTrees} (TC for short). Let us stress the differences between these two problems:
\begin{itemize}
  \item In TC, two-qubit operations are either realisable on the QPU topology or not. In QCD, the operations are all realisable, but are either local (cheap) or nonlocal (requiring expensive ebit communication). The outcome of TC is a circuit that only uses realisable operations, while QCD's outcome uses both local and nonlocal operations. 
  \item TC attempts to find the shallowest equivalent circuit, whereas QCD only focuses on reducing the number of nonlocal (communication) operations. The most efficient circuit communication-wise may not be the smallest depth-wise.
  \item TC is a \emph{local} optimisation problem: for each unrealisable operation, the optimal way to simulate it must be found, which may depend on the way the operations in its immediate neighbourhood are implemented. In QCD, rather than optimising each nonlocal operation separately, the focus is on the \emph{global} interaction between qubits, as we need to group highly interacting qubits together, so that communication across QPUs is minimal.
\end{itemize}
The problem of distribution of quantum circuits has not received much attention in the literature. Some authors have previously proposed a solution using standard graph partitioning~\cite{PrevSol}. However, in that work, an additional exhaustive search is applied to decide how each two-qubit quantum gate should be implemented. This increases the runtime  exponentially compared to the input size, making it futile in practice. The algorithm we propose in Section~\ref{sec:algorithm} encodes all possible choices of distribution  in a hypergraph, and so our optimisation procedure relies solely on a hypergraph partitioner. The latter programs have been extensively studied and perfected in the computer science literature to perform efficiently even for large inputs~\cite{KaHyPar, PaToH}. Unlike the former work~\cite{PrevSol}, our approach may distribute circuits across any number of QPUs, thus answering an open problem proposed by the previous authors.


The idea of treating quantum circuits as graphs has been previously used in the literature. For instance, graphs have been employed to represent the causal structure of circuits for applications such as the recycling of circuit wires~\cite{WireRecycling}. Certain results on classical simulation of quantum circuits have been found through graph-theoretic approaches~\cite{ClassicalSimulation}.

\section{Automated circuit distribution}
\label{sec:algorithm}

In this section we describe how the problem of quantum circuit distribution can be solved using hypergraph partitioning. We discuss some aspects of its practical implementation. Further technical details are given in Appendices A and B.

\subsection{Reduction to hypergraph partitioning}
\label{sec:reduction}

A hypergraph consists of a set $V$ of vertices and a set $H$ of hyperedges, each hyperedge being defined as the subset of vertices it connects \(\forall h\in H,\ h \subseteq V\). The hypergraph partitioning problem has as input a hypergraph $(V,H)$, a parameter $k$ giving the number of blocks (sub-hypergraphs) we wish to partition the hypergraph into, and a parameter $\omega$ known as the load-imbalance tolerance. The output is a labelling $f\colon V \to \{1,2 \dots k\}$ of vertices to blocks, satisfying the following two criteria:
\begin{itemize}
  \item \emph{load-balance}: for all $i=1,2,\ldots,k$: 
  \begin{equation}
    \label{eq:balance}
    \lvert \{v \in V \mid f(v) = i\} \rvert \, < \, (1+\omega)\frac{\lvert V \rvert}{k} 
  \end{equation}
  which implies that a valid labelling \(f\) must assign roughly the same amount of vertices to each block, with $\omega$ acting as a tolerance parameter.
  \item \emph{minimal number of cuts}: given a way to assign a score $\chi_g \in \mathbb{N}$ to a labelling $g \colon V \to \{1,2,\ldots,k\}$, the score of the output $\chi_f$ should be the lowest possible: $\forall g,\ \chi_f \leq \chi_g$.
  The score may be calculated in several ways, corresponding to variations of the hypergraph partitioning problem. We use $\chi_g = \sum_{h \in H} \ \lambda_g(h)$, where
  \begin{equation}
    \label{eq:cuts}
    \lambda_g(h) = \lvert \{i \in \mathbb{N} \mid \exists v \in h,\ g(v)=i \} \rvert - 1.
  \end{equation}
  Equation (\ref{eq:cuts}) calculates, for each hyperedge $h\in H$, the number of different blocks its vertices have been assigned to, and substracts one in order to obtain the number of times the hyperedge is cut; e.g.\ if all the vertices of $h$ are in the same block, \(\lambda(h) = 0\).
\end{itemize}

We reduce the problem of efficiently distributing quantum circuits to the problem of hypergraph partitioning along the following intuition:
\begin{center}
\begin{tabular}{c|c}
\emph{\  Hypergraph partitioning\ } & \emph{\ Circuit distribution\ } \\
\hline
vertices & wires (qubits) \\
hyperedges & groups of CZs \\
partition & distribution \\
blocks & QPUs \\
load-balance \eqref{eq:balance} & load-balance \\
fewest cuts \eqref{eq:cuts} & fewest ebits used \\
\end{tabular}
\end{center}

The algorithm in Figure~\ref{code:buildHyp} encodes all information about how the circuit's CZ gates may be grouped together (i.e.\ when they may share the same ebit, see Figure~\ref{fig:nonlocal}) by representing such groups as a single hyperedge. The algorithm runs in time linear $\mathcal{O}(n)$ in the number $n$ of gates of the input circuit. Figure~\ref{fig:process} shows an example execution. Each vertex in the hypergraph corresponds to either a wire or a CZ gate; we will refer to them as wire-vertices and CZ-vertices respectively. The following theorem is the key insight that makes our approach successful. 

\begin{figure}[b]
\begin{algorithm}
input: circuit
output: (V,H)
begin
  V $\gets$ $\varnothing$
  H $\gets$ $\varnothing$
  foreach wire in circuit do
    V $\gets$ V $\cup$ {wire}
    hedge $\gets$ {wire}
    foreach gate in wire do
      if $isCZ$(gate) then
        V $\gets$ V $\cup$ {gate}
        hedge $\gets$ hedge $\cup$ {gate}
      else
        H $\gets$ H $\uplus$ {hedge}
        hedge $\gets$ {wire}
    H $\gets$ H $\uplus$ {hedge}
end
\end{algorithm}
\caption{Pseudocode for the algorithm that translates a quantum circuit into a hypergraph containing all relevant qubit interaction information.}
\label{code:buildHyp}
\end{figure}

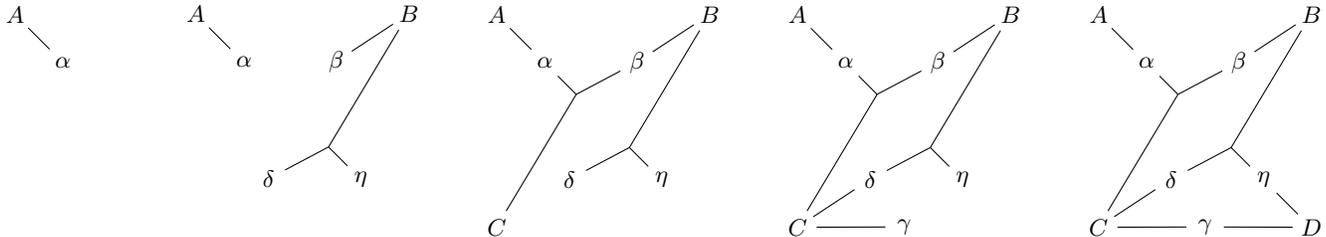
\begin{figure*}
  \[\begin{tikzpicture}
    \node (1) at (-2.5,1.1) {
    \begin{tikzpicture}
      \coordinate (A) at (135:20mm);
      \coordinate (a) at (135:11mm);
      \draw (A) -- (a);
      \node[circle, right=-2.5mm of A, fill=white, inner sep=0pt, minimum size=5mm] {\(A\)};
      \node[circle, right=-2.5mm of a, fill=white, inner sep=0pt, minimum size=5mm] {\(\alpha\)};
    \end{tikzpicture}};
    \node (2) at (1,.325) {
    \begin{tikzpicture}
      \coordinate (auxB) at (315:5mm);
      \coordinate (A) at (135:20mm);
      \coordinate (B) at (45:20mm);
      \coordinate (a) at (135:11mm);
      \coordinate (b) at (60:9mm);
      \coordinate (d) at (240:9mm);
      \coordinate (e) at (315:11mm);
      \draw (auxB) -- (B);
      \draw (auxB) -- (d);
      \draw (auxB) -- (e);
      \draw (B) -- (b);
      \draw (A) -- (a);
      \node[circle, right=-2.5mm of A, fill=white, inner sep=0pt, minimum size=5mm] {\(A\)};
      \node[circle, right=-2.5mm of B, fill=white, inner sep=0pt, minimum size=5mm] {\(B\)};
      \node[circle, right=-2.5mm of a, fill=white, inner sep=0pt, minimum size=5mm] {\(\alpha\)};
      \node[circle, right=-2.5mm of b, fill=white, inner sep=0pt, minimum size=5mm] {\(\beta\)};
      \node[circle, right=-2.5mm of d, fill=white, inner sep=0pt, minimum size=5mm] {\(\delta\)};
      \node[circle, right=-2.5mm of e, fill=white, inner sep=0pt, minimum size=5mm] {\(\eta\)};
    \end{tikzpicture}};
    \node (3) at (5,0) {
    \begin{tikzpicture}
      \coordinate (auxC) at (135:5mm);
      \coordinate (auxB) at (315:5mm);
      \coordinate (A) at (135:20mm);
      \coordinate (B) at (45:20mm);
      \coordinate (C) at (225:20mm);
      \coordinate (a) at (135:11mm);
      \coordinate (b) at (60:9mm);
      \coordinate (d) at (240:9mm);
      \coordinate (e) at (315:11mm);
      \draw (auxC) -- (C);
      \draw (auxC) -- (a);
      \draw (auxC) -- (b);
      \draw (auxB) -- (B);
      \draw (auxB) -- (d);
      \draw (auxB) -- (e);
      \draw (B) -- (b);
      \draw (A) -- (a);
      \node[circle, right=-2.5mm of A, fill=white, inner sep=0pt, minimum size=5mm] {\(A\)};
      \node[circle, right=-2.5mm of B, fill=white, inner sep=0pt, minimum size=5mm] {\(B\)};
      \node[circle, right=-2.5mm of C, fill=white, inner sep=0pt, minimum size=5mm] {\(C\)};
      \node[circle, right=-2.5mm of a, fill=white, inner sep=0pt, minimum size=5mm] {\(\alpha\)};
      \node[circle, right=-2.5mm of b, fill=white, inner sep=0pt, minimum size=5mm] {\(\beta\)};
      \node[circle, right=-2.5mm of d, fill=white, inner sep=0pt, minimum size=5mm] {\(\delta\)};
      \node[circle, right=-2.5mm of e, fill=white, inner sep=0pt, minimum size=5mm] {\(\eta\)};
    \end{tikzpicture}};
    \node (4) at (9,0) {
    \begin{tikzpicture}
      \coordinate (auxC) at (135:5mm);
      \coordinate (auxB) at (315:5mm);
      \coordinate (A) at (135:20mm);
      \coordinate (B) at (45:20mm);
      \coordinate (C) at (225:20mm);
      \coordinate (a) at (135:11mm);
      \coordinate (b) at (60:9mm);
      \coordinate (c) at (270:14.1mm);
      \coordinate (d) at (240:9mm);
      \coordinate (e) at (315:11mm);
      \draw (auxC) -- (C);
      \draw (auxC) -- (a);
      \draw (auxC) -- (b);
      \draw (auxB) -- (B);
      \draw (auxB) -- (d);
      \draw (auxB) -- (e);
      \draw (C) -- (d);
      \draw (B) -- (b);
      \draw (A) -- (a);
      \draw (C) -- (c);
      \node[circle, right=-2.5mm of A, fill=white, inner sep=0pt, minimum size=5mm] {\(A\)};
      \node[circle, right=-2.5mm of B, fill=white, inner sep=0pt, minimum size=5mm] {\(B\)};
      \node[circle, right=-2.5mm of C, fill=white, inner sep=0pt, minimum size=5mm] {\(C\)};
      \node[circle, right=-2.5mm of a, fill=white, inner sep=0pt, minimum size=5mm] {\(\alpha\)};
      \node[circle, right=-2.5mm of b, fill=white, inner sep=0pt, minimum size=5mm] {\(\beta\)};
      \node[circle, right=-2.5mm of c, fill=white, inner sep=0pt, minimum size=5mm] {\(\gamma\)};
      \node[circle, right=-2.5mm of d, fill=white, inner sep=0pt, minimum size=5mm] {\(\delta\)};
      \node[circle, right=-2.5mm of e, fill=white, inner sep=0pt, minimum size=5mm] {\(\eta\)};
    \end{tikzpicture}};
    \node (5) at (13,0) {
    \begin{tikzpicture}
      \coordinate (auxC) at (135:5mm);
      \coordinate (auxB) at (315:5mm);
      \coordinate (A) at (135:20mm);
      \coordinate (B) at (45:20mm);
      \coordinate (C) at (225:20mm);
      \coordinate (D) at (315:20mm);
      \coordinate (a) at (135:11mm);
      \coordinate (b) at (60:9mm);
      \coordinate (c) at (270:14.1mm);
      \coordinate (d) at (240:9mm);
      \coordinate (e) at (315:11mm);
      \draw (auxC) -- (C);
      \draw (auxC) -- (a);
      \draw (auxC) -- (b);
      \draw (auxB) -- (B);
      \draw (auxB) -- (d);
      \draw (auxB) -- (e);
      \draw (C) -- (d);
      \draw (D) -- (e);
      \draw (B) -- (b);
      \draw (A) -- (a);
      \draw (D) -- (c);
      \draw (C) -- (c);
      \node[circle, right=-2.5mm of A, fill=white, inner sep=0pt, minimum size=5mm] {\(A\)};
      \node[circle, right=-2.5mm of B, fill=white, inner sep=0pt, minimum size=5mm] {\(B\)};
      \node[circle, right=-2.5mm of C, fill=white, inner sep=0pt, minimum size=5mm] {\(C\)};
      \node[circle, right=-2.5mm of D, fill=white, inner sep=0pt, minimum size=5mm] {\(D\)};
      \node[circle, right=-2.5mm of a, fill=white, inner sep=0pt, minimum size=5mm] {\(\alpha\)};
      \node[circle, right=-2.5mm of b, fill=white, inner sep=0pt, minimum size=5mm] {\(\beta\)};
      \node[circle, right=-2.5mm of c, fill=white, inner sep=0pt, minimum size=5mm] {\(\gamma\)};
      \node[circle, right=-2.5mm of d, fill=white, inner sep=0pt, minimum size=5mm] {\(\delta\)};
      \node[circle, right=-2.5mm of e, fill=white, inner sep=0pt, minimum size=5mm] {\(\eta\)};
    \end{tikzpicture}};
  \end{tikzpicture}\]
  \caption{Step by step execution of the algorithm in Figure~\ref{code:buildHyp} with input the quantum circuit of Figure~\ref{fig:circuit}. Each hyperedge is represented as a collection of line segments that all meet at one end, while their other ends reach each of the hyperedge's vertices. Greek letters represent CZ-vertices, latin letters represent wire-vertices. \label{fig:process}}
\end{figure*}

\vspace{1ex}
\textbf{Theorem} \emph{Given a circuit, each of its possible distributed implementations (without altering the gateset or the gate order) corresponds to a unique partition of its hypergraph (given by Figure~\ref{code:buildHyp}) whose number of cuts equals the number of ebits required.}
\vspace{1ex}

The theorem implies that we may use third-party hypergraph partitioners to produce circuit distributions with low ebit count. We now explain the intuition behind the theorem. First, observe that any distribution is described by a hypergraph partition: assigning a wire-vertex to a block indicates in which QPU the corresponding wire is allocated. Similarly, assigning a CZ-vertex to a block determines which QPU will perform the CZ operation. Accordingly, the CZ gate will be local or require communication (i.e.\ ebits) to access its target wires. Notice that in Figure~\ref{fig:process} each hyperedge connects a wire-vertex with multiple CZ-vertices: it represents all the locations where the wire's state is required. The number of cuts of a given hyperedge corresponds to the number of extra blocks it reaches~\eqref{eq:cuts}, and for each of them an ebit is needed so the wire's state is accessible. Therefore, the number of cuts corresponds precisely to the number of ebits. Appendix A gives a detailed proof of the theorem.

To build the distributed circuit, add a cat-entangler and cat-disentangler for each cut, and then allocate all CZ gates to their corresponding QPU, connecting the relevant wires and ebit halves. This translation takes $\mathcal{O}(\textit{cuts}+\textit{gates})$ steps. However, by construction of the hypergraph, we know that $\textit{cuts} \leq 2\, \textit{gates}$, and thus this transformation takes time $\mathcal{O}(n)$, i.e.\ linear in the number $n$ of gates from the original circuit. 

\subsection{Implementation}
\label{sec:implementation}

Reducing the problem to hypergraph partitioning lets us use third-party solvers such as KaHyPar~\cite{KaHyPar}.
We implemented this approach in the quantum circuit description language Quipper~\cite{quipperonline}; the code is available at~\cite{pablogithub}.

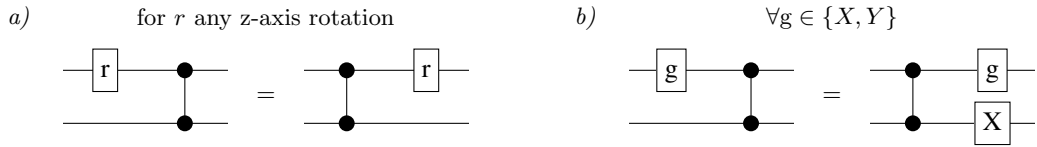
\begin{figure*}
\hspace*{-6mm}
\begin{tikzpicture}
  \node (A) {
    \begin{tikzpicture}
      \node[inner sep=0pt] (c1) at (0,0) {\includegraphics[scale=2]{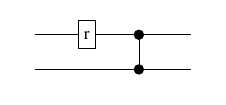}};       
      \node[right=-13.9mm of c1.east, inner sep=0pt] (c2) {\includegraphics[scale=2]{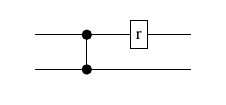}};
      \node[right=-12mm of c1.east, rectangle, fill=white, minimum size=10mm] (eq) {\(=\)};  
      \node[right=7mm of c1.west, rectangle,fill=white,minimum width=5mm, minimum height=10mm] {};
      \node[left=7mm of c2.east, rectangle,fill=white,minimum width=5mm, minimum height=10mm] {};
    \end{tikzpicture}
  };
  \node[above=-4mm of A] (textA) {for $r$ any z-axis rotation};
  \node[above left=-4mm and -10mm of A] (a) {\emph{a)}};
  \node [right=-5mm of A] (B) {
    \begin{tikzpicture}
      \node[inner sep=0pt] (c1) at (0,0) {\includegraphics[scale=2]{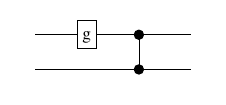}};       
      \node[right=-13.9mm of c1.east, inner sep=0pt] (c2) {\includegraphics[scale=2]{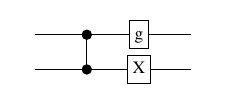}};
      \node[right=-12mm of c1.east, rectangle, fill=white, minimum size=10mm] (eq) {\(=\)};  
      \node[right=7mm of c1.west, rectangle,fill=white,minimum width=5mm, minimum height=10mm] {};
      \node[left=7mm of c2.east, rectangle,fill=white,minimum width=5mm, minimum height=10mm] {};
    \end{tikzpicture}
  };
  \node[above=-4mm of B] (textB) {\(\forall \text{g}\in\{X,Y\}\)};
  \node[above left=-4mm and -10mm of B] (b) {\emph{b)}};
\end{tikzpicture}
\vspace{-3ex}
\caption{Well-known cases where a 1-qubit gate can be pushed through a CZ gate. In case \emph{b)} a byproduct gate is created. These byproduct gates can in turn be pushed through other CZs.}
\label{fig:pullRules}
\end{figure*}

Apart from extracting a hypergraph out of the input circuit (Figure~\ref{code:buildHyp}), and building the distributed circuit from the resulting partition, we include some additional pre-processing and post-processing phases:

\begin{itemize}
  \item \emph{Pre-processing 1}: transform the input circuit into an equivalent one using only one-qubit gates and CZ gates; Quipper provides specialised functionality to do so. 
  \item \emph{Pre-processing 2}: use the well-known rules from Figure~\ref{fig:pullRules} to reorder CZ gates and some 1-qubit gates, pulling CZ gates as early in the circuit as possible. This brings CZ gates closer together, letting our algorithm implement larger groups of nonlocal CZ gates using a single ebit. As shown in Figure~\ref{fig:pullRules}, doing so may create new 1-qubit gates, namely Pauli X gates. Using the same rules, these byproduct gates can be pushed to the end of the circuit, where they will cancel out with other byproduct gates, so the overhead is bounded by at most a single pair of extra Pauli (X and Z) gates per wire.
  \item \emph{Pre-processing 3}: in many circuits, the main group of qubits that another qubit interacts with varies between the different stages of the circuit. Then, if we were to use the hypergraph of the whole circuit, the different connectivities of each stage would be confounded, preventing the hypergraph partitioner from properly optimising them. To account for this, we first run a procedure that detects significant changes in the circuit's qubit connectivity and  splits the circuit into multiple segments accordingly. Each of these segments is then distributed using the approach presented in Section~\ref{sec:reduction}. Appendix B details this extra pre-processing procedure.
  \item \emph{Post-processing}: reduce the required ebit storage space by garbage management while building the distributed circuit: immediately after performing the last CZ gate of a group that involves an ebit, apply its cat-disentangler. This destroys the ebit so its space can be reused to store a newly created ebit.
\end{itemize}  

\section{Results}
\label{sec:results}

Our algorithm was evaluated on five quantum circuits provided by Quipper's library~\cite{quipperonline}. These circuits implement algorithms that have been discussed in the literature as examples where quantum computers achieve computational speedup. Their default configuration was used unless stated otherwise:
\begin{itemize}
  \item \emph{Boolean formula (BF)}~\cite{BFWalk}: the circuit fragment implementing the quantum walk, the core of the algorithm;
  \item \emph{Binary welded tree (BWT)}~\cite{BWT}: tree height set to \(200\) (from \(5\) by default);
  \item \emph{Ground state estimation (GSE)}~\cite{GSE}: number of precision qubits increased to \(150\) (from \(3\) by default);
  \item \emph{Unique shortest vector (USV-R)}~\cite{USV}: the subproblem called `R', with lattice dimension \(3\) (from \(5\) by default);
  \item \emph{Quantum Fourier transform (QFT)}~\cite{nielsenchuang}: using \(200\) qubits.
\end{itemize}

\begin{table}
\caption{Original number of qubits and CZ gates of each of the circuits. We distributed each of them across \(k\) different QPUs, with \(4 \leq k \leq 16\). Columns `Ebit space overhead' and `Time' show the worst value among these distributions.}
\begin{tabular}{|c|cccc|}
\hline
Circuit & Qubits &  \hspace{0.5em} CZ gates \hspace{0.5em} & Time & \begin{tabular}{c}Ebit space \\ overhead\end{tabular} \\
\hline
BF & 105 & 25,590 & 23.00s & 4.8\% \\
BWT & 614 & 261,456 & 389.39s & 1.1\% \\
GSE & 156 & 237,750 & 307.75s & 2.6\% \\
USV-R & 842 & 377,695 & 282.32s & 2.4\% \\
QFT & 201 & 199,000 & 327.48s & 4.0\% \\
\hline
\end{tabular}
\label{tab:perf}
\end{table}

Table~\ref{tab:perf} shows the number of qubits and CZ gates of each circuit. These circuits require more qubits than the number a single near-term QPU can handle, making them meaningful examples on which to evaluate the distribution approach proposed in this paper. The times shown in the table indicate how long it took to obtain the distributed circuit using our implementation (available at~\cite{pablogithub}), running it on a standard desktop computer. The fact that circuits of this size can be distributed in a few minutes shows that our approach is useful in practice. 

The last column from Table~\ref{tab:perf} shows the percentage of extra quantum memory required to store the ebit halves used for communication. The proportion is calculated by counting the maximum number of ebit halves stored simultaneously, and dividing it by the number of qubits in the original circuit. This overhead was considerably reduced from previous versions of our approach by limiting the number of gates allowed to be applied between two nonlocal CZ gates sharing an ebit, so that the corresponding ebit does not need to be stored over a long period of time. 

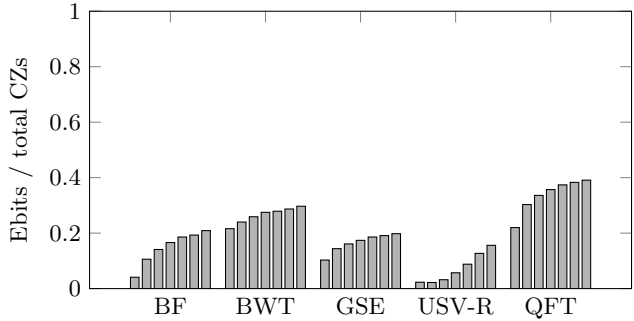
\begin{figure}
\begin{tikzpicture}
  \pgfplotsset{ybar stacked}
  \begin{axis}[
      ylabel=Ebits / total CZs,
      ylabel near ticks,
      height=150pt,
      width=250pt,
      bar width=3.25pt,      
      xtick={4,12,20,28,36},
      xticklabels={BF,BWT,GSE,USV-R,QFT},
      ymin=0,
      ymax=1,
      legend pos=north west,
      legend style={font=\scriptsize},
      legend cell align={left}]
    \addplot [fill=black!30,] table [x=X, y=EbitsOverCZs]  {ebitEff.dat};
  \end{axis}
\end{tikzpicture}
\caption{Each bar shows the proportion of ebits required over the total number of CZs. For each circuit and from left to right, the bars correspond to distributing the circuit across \(4\), \(6\), \(8\), \(10\), \(12\), \(14\) and \(16\) QPUs with an equal number of qubits allocated to each.}
\label{fig:ebitEff}
\end{figure}

Figure~\ref{fig:ebitEff} shows the proportion of ebits required when each circuit was distributed. In all cases, over \(60\%\) of the CZ gates could be implemented either locally or `for free' using already existing ebits. Naturally, as the number of QPUs used to distribute the circuit increases, more communication is required among them and a larger number of ebits is used. In practice, the number of QPUs each circuit should be distributed across will be determined by the circuit size; for instance, GSE may be distributed across \(8\) QPUs, each managing \(20\) qubits.

\begin{figure}
\begin{tikzpicture}
  \pgfplotsset{ybar stacked}
  \begin{axis}[
      ylabel=Ebits / total CZs,
      ylabel near ticks,
      height=150pt,
      width=250pt,
      bar width=10pt,      
      xtick={3,11,19,27,35},
      xticklabels={BF,BWT,GSE,USV-R,QFT},
      ymin=0,
      ymax=0.6,
      legend pos=north west,
      legend style={font=\scriptsize},
      legend cell align={left}]
    \addplot [fill=black!30] table [x=X, y=EbitsOverCZs]  {hypImprove.dat};
    \addplot [pattern=north west lines, pattern color=black!90] table [x=X, y=ExtraNonlocal]  {hypImprove.dat};
    \addlegendentry{Ebits required};
    \addlegendentry{Ebits saved};
  \end{axis}
\end{tikzpicture}
\caption{The gray bar indicates the proportion of ebits required in the distribution found using hypergraph partitioning. The bar on top indicates the extra ebits if standard graphs were used instead. For all circuits, the data corresponds to distributing them across \(8\) QPUs.}
\label{fig:hypImprove}
\end{figure}
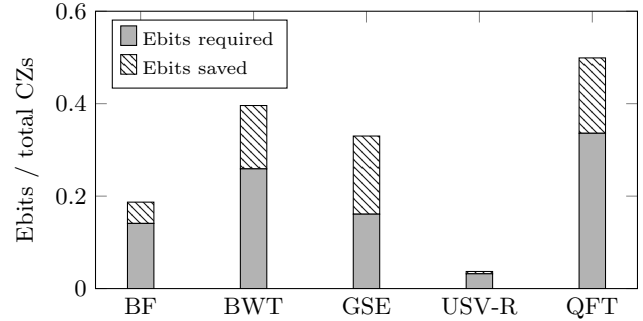

To put the quality of these results into perspective, Figure~\ref{fig:hypImprove} compares our approach with a simplified version using standard graph partitioning instead of hypergraph partitioning.  The use of hypergraphs is the main contribution of our approach: in previous works~\cite{PrevSol} the optimisation of the number of nonlocal gates that are implemented `for free' was achieved by exhaustively exploring the space of all possible configurations, which is exponential in size and therefore not workable in practice. Thanks to state-of-the-art hypergraph partitioners, this optimisation can be achieved in a practical amount of time by letting heuristics guide the search, instead of trying each possible configuration. The proportion of `Ebits saved' as labelled in Figure~\ref{fig:hypImprove} corresponds to the number of CZ gates that were implemented `for free' using our approach.  In some cases, such as GSE, we are saving approximately half of the number of ebits; in other cases, such as USV-R, the improvement is almost unnoticeable because most CZ gates are already implemented locally.

\subsection*{Bonus: Distributing CCZ gates}

For certain DQAs, it has been discussed that local Toffoli gates could be computed at approximately the same cost as a local CZ gate~\cite{HierarchichalIonTrap}. Toffoli gates are three-qubit gates extensively used in quantum circuits and, in most architectures, they are implemented by decomposing each one into multiple one-qubit gates and \(6\) CNOT gates~\cite{ToffoliDecomp}. Interestingly, we can easily adapt our approach to distribute CCZ gates, which are locally equivalent to the Toffoli gate by applying one-qubit Hadamard gates. 

To extend our approach to this setting, it suffices to realise that the approach from Figure~\ref{fig:nonlocal} can be used to implement groups of nonlocal CCZ gates together with CZ gates; for instance, if CZ gate \(\alpha\) from Figure~\ref{fig:nonlocal} is replaced by a CCZ gate acting on the three qubits, the same cat-entangler and cat-disentangler allow us to implement both the CCZ gate and CZ gate \(\beta\) using a single ebit. After all, this cat-entangler and this cat-disentangler are compatible with the qubit basis the CCZ gate acts upon (which is the same as CZ). If a CCZ gate has each of its three wires allocated to different QPUs, two ebits will be required to implement it, so that the quantum information in each of the wires can be accessed by the QPU where the CCZ is actually applied. When building our hypergraph from the circuit (as in Figure~\ref{code:buildHyp}), CCZ-vertices are created in the same way CZ-vertices are, but in this case each of them would be reached by three hyperedges: one per wire the gate acts upon. The rest of the approach works exactly the same as described before, consuming an ebit whenever a cut appears in the hypergraph partition, regardless of whether it reaches a CCZ-vertex.

\begin{figure}
\begin{tikzpicture}
  \pgfplotsset{ybar stacked}
  \begin{axis}[
      ylabel=Ebits / total CZs,
      ylabel near ticks,
      height=150pt,
      width=250pt,
      bar width=10pt,      
      xtick={3,11,19,27,35},
      xticklabels={BF,BWT,GSE,USV-R,QFT},
      ymin=0,
      ymax=0.6,
      legend pos=north west,
      legend style={font=\scriptsize},
      legend cell align={left}]
    \addplot [fill=black!30] table [x=X, y=CC_EbitsOverCZs]  {cczImprove.dat};
    \addplot [pattern=north west lines, pattern color=black!90] table [x=X, y=ExtraEbits]  {cczImprove.dat};
    \addlegendentry{Ebits required};
    \addlegendentry{Ebits saved};
  \end{axis}
\end{tikzpicture}
\caption{The gray bar indicates the proportion of ebits required when using the extension where CCZ gates are distributed. The bar on top indicates the extra ebits if the CCZ gates are decomposed into CZ gates instead. In both cases, hypergraph partitioning is used. For all circuits, the data corresponds to distributing them across \(8\) QPUs.}
\label{fig:cczImprove}
\end{figure}
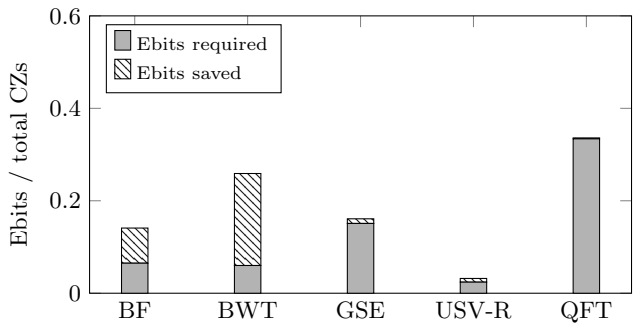

If an architecture allows implementing local CCZ gates directly, this extended approach would yield distributed circuits requiring fewer ebits: the three qubits of a CCZ only interact once, whereas if the CCZ gate is implemented using \(6\) CZ gates, the communication required is increased. Figure~\ref{fig:cczImprove} shows that distributing CCZ gates saves a remarkable proportion of ebits for circuits using \(n\)-qubit gates (with \(n > 2\)) extensively: BF and BWT. Naturally, this extension does not change the result when only two-qubit gates are applied between qubits, such as in QFT.

\section{Discussion and further work}
\label{sec:discussion}

The lemma from Appendix A states that, in a similar way we may use hypergraph partitioning to solve the problem of quantum circuit distribution, the other way around is also feasible. This implies that if someone could devise an optimisation procedure that beats our distribution approach (i.e.\ gives better results and takes less time), we could immediately convert such a procedure into a hypergraph partitioner that beats KaHyPar~\cite{KaHyPar}, the state-of-art hypergraph partitioner we used. Considering that hypergraph partitioning has been extensively studied by experts in algorithm design~\cite{KaHyPar,PaToH,NP-hard}, it is unlikely that a dramatically better approach to quantum circuit distribution exists unless some of the constraints we imposed are lifted. These constraints are described below; they constitute the open problems that should be addressed in order to reduce the communication cost of distribution even further:

\begin{itemize}
  \item \emph{Gateset}. Our chosen gateset contained every one-qubit gate and a single two-qubit gate: the CZ gate. Gatesets where other multi-qubit gates are allowed may bring better results. Our approach is easily adapted to use other gatesets, as shown in Section~\ref{sec:results} where the CCZ gate was included in our gateset. The question of which gateset is best for distribution is left as an open problem, and we point out that this may be architecture-dependent.

  \item \emph{Rearranging multi-qubit gates}. The procedure labelled pre-processing 2 in Section~\ref{sec:implementation} rearranges one-qubit gates in the circuit to create larger groupings of CZ gates, which can reduce the number of ebits required to distribute the circuit. It is likely that, by rearranging multi-qubit gates, the connectivity of the circuit may be changed in a way that favours distribution.
\end{itemize}

It is important to stress that hypergraph partitioners are not expected to provide the optimal partition of the input hypergraph: that problem is intractable on a classical computer (namely, it is NP-hard~\cite{NP-hard}). Instead, we work with close to optimal solutions that can be found efficiently by classical computers. The results discussed in Section~\ref{sec:results} were obtained using such suboptimal partitions. These allow us to reduce the communication cost of distributing circuits, and thus help us compute problems that are not tractable in classical computers (not even suboptimally) and whose quantum circuits would require more qubits than the number a near-term QPU can handle.

Apart from the restricted number of qubits a QPU can manage, there is another fundamental limit to scalability we have overlooked up to this point: the short lifespan of qubits due to decoherence. NISQ computers will only be able to store and manipulate quantum information for a short period of time, which means we should not expect to be able to execute more than \(1000\) consecutive two-qubit gates~\cite{NISQ}. This means that optimising the depth of the circuit (i.e.\ reducing the largest chain of consecutive gates) is considered essential. There are many different methods that reduce the depth of circuits~\cite{CQCReduction,TReduction}, and we propose these should be used to optimise the input circuit before distributing it with our approach. An interesting line of research is exploiting how parallelism may be employed to further reduce the circuit depth: if a circuit is distributed across different QPUs, the QPUs may perform simultaneous computations, reducing the total time the quantum information needs to be coherently stored.

We have seen that distributed quantum architectures~\cite{HierarchichalIonTrap, HierarchichalQuantumDot} have been proposed as a feasible approach to increase the size of quantum computers.
Circuits that are too large to be performed in near-term quantum processing units may be run on distributed quantum architectures at the cost of quantum communication.
We have presented an automated method for distributing quantum circuits across multiple agents, minimising the quantum communication between them.
In this last section we have discussed the limitations of our approach and pointed out further lines of research that would improve it. 
Our approach was evaluated favourably on five test circuits of interest to the quantum computing literature. These circuits are too large to fit in a single near-term QPU and thus need to be distributed in order to be implemented. 

\bibliography{Bibliography}

\begin{thebibliography}{47}%
\makeatletter
\providecommand \@ifxundefined [1]{%
 \@ifx{#1\undefined}
}%
\providecommand \@ifnum [1]{%
 \ifnum #1\expandafter \@firstoftwo
 \else \expandafter \@secondoftwo
 \fi
}%
\providecommand \@ifx [1]{%
 \ifx #1\expandafter \@firstoftwo
 \else \expandafter \@secondoftwo
 \fi
}%
\providecommand \natexlab [1]{#1}%
\providecommand \enquote  [1]{``#1''}%
\providecommand \bibnamefont  [1]{#1}%
\providecommand \bibfnamefont [1]{#1}%
\providecommand \citenamefont [1]{#1}%
\providecommand \href@noop [0]{\@secondoftwo}%
\providecommand \href [0]{\begingroup \@sanitize@url \@href}%
\providecommand \@href[1]{\@@startlink{#1}\@@href}%
\providecommand \@@href[1]{\endgroup#1\@@endlink}%
\providecommand \@sanitize@url [0]{\catcode `\\12\catcode `\$12\catcode
  `\&12\catcode `\#12\catcode `\^12\catcode `\_12\catcode `\%12\relax}%
\providecommand \@@startlink[1]{}%
\providecommand \@@endlink[0]{}%
\providecommand \url  [0]{\begingroup\@sanitize@url \@url }%
\providecommand \@url [1]{\endgroup\@href {#1}{\urlprefix }}%
\providecommand \urlprefix  [0]{URL }%
\providecommand \Eprint [0]{\href }%
\providecommand \doibase [0]{https://doi.org/}%
\providecommand \selectlanguage [0]{\@gobble}%
\providecommand \bibinfo  [0]{\@secondoftwo}%
\providecommand \bibfield  [0]{\@secondoftwo}%
\providecommand \translation [1]{[#1]}%
\providecommand \BibitemOpen [0]{}%
\providecommand \bibitemStop [0]{}%
\providecommand \bibitemNoStop [0]{.\EOS\space}%
\providecommand \EOS [0]{\spacefactor3000\relax}%
\providecommand \BibitemShut  [1]{\csname bibitem#1\endcsname}%
\let\auto@bib@innerbib\@empty
\bibitem [{\citenamefont {Feynman}(1982)}]{feynman}%
  \BibitemOpen
  \bibfield  {author} {\bibinfo {author} {\bibfnamefont {R.~P.}\ \bibnamefont
  {Feynman}},\ }\bibfield  {title} {\bibinfo {title} {Simulating physics with
  computers},\ }\href {https://doi.org/10.1007/BF02650179} {\bibfield
  {journal} {\bibinfo  {journal} {Int. J. Th. Phys.}\ }\textbf {\bibinfo
  {volume} {21}},\ \bibinfo {pages} {467} (\bibinfo {year} {1982})}\BibitemShut
  {NoStop}%
\bibitem [{\citenamefont {Nielsen}\ and\ \citenamefont
  {Chuang}(2000)}]{nielsenchuang}%
  \BibitemOpen
  \bibfield  {author} {\bibinfo {author} {\bibfnamefont {M.~J.}\ \bibnamefont
  {Nielsen}}\ and\ \bibinfo {author} {\bibfnamefont {I.~L.}\ \bibnamefont
  {Chuang}},\ }\href@noop {} {\emph {\bibinfo {title} {Quantum computation and
  quantum information}}}\ (\bibinfo  {publisher} {Cambridge University Press},\
  \bibinfo {year} {2000})\BibitemShut {NoStop}%
\bibitem [{\citenamefont {Lloyd}(1996)}]{lloyd:simulators}%
  \BibitemOpen
  \bibfield  {author} {\bibinfo {author} {\bibfnamefont {S.}~\bibnamefont
  {Lloyd}},\ }\bibfield  {title} {\bibinfo {title} {Universal quantum
  simulators},\ }\href {https://doi.org/10.1126/science.273.5278.1073}
  {\bibfield  {journal} {\bibinfo  {journal} {Science}\ }\textbf {\bibinfo
  {volume} {273}},\ \bibinfo {pages} {1073} (\bibinfo {year}
  {1996})}\BibitemShut {NoStop}%
\bibitem [{\citenamefont {Shor}(1999)}]{Shor}%
  \BibitemOpen
  \bibfield  {author} {\bibinfo {author} {\bibfnamefont {P.}~\bibnamefont
  {Shor}},\ }\bibfield  {title} {\bibinfo {title} {Polynomial-time algorithms
  for prime factorization and discrete logarithms on a quantum computer},\
  }\href {https://doi.org/10.1137/S0036144598347011} {\bibfield  {journal}
  {\bibinfo  {journal} {SIAM Review}\ }\textbf {\bibinfo {volume} {41}},\
  \bibinfo {pages} {303} (\bibinfo {year} {1999})}\BibitemShut {NoStop}%
\bibitem [{\citenamefont {Grover}(1996)}]{Grover}%
  \BibitemOpen
  \bibfield  {author} {\bibinfo {author} {\bibfnamefont {L.~K.}\ \bibnamefont
  {Grover}},\ }\bibfield  {title} {\bibinfo {title} {A fast quantum mechanical
  algorithm for database search},\ }in\ \href@noop {} {\emph {\bibinfo
  {booktitle} {Annual ACM symposium on theory of computing}}}\ (\bibinfo
  {publisher} {ACM},\ \bibinfo {year} {1996})\ pp.\ \bibinfo {pages}
  {212--219}\BibitemShut {NoStop}%
\bibitem [{\citenamefont {Harrow}\ \emph {et~al.}(2009)\citenamefont {Harrow},
  \citenamefont {Hassidim},\ and\ \citenamefont
  {Lloyd}}]{harrowhassadimlloyd:linear}%
  \BibitemOpen
  \bibfield  {author} {\bibinfo {author} {\bibfnamefont {A.~W.}\ \bibnamefont
  {Harrow}}, \bibinfo {author} {\bibfnamefont {A.}~\bibnamefont {Hassidim}},\
  and\ \bibinfo {author} {\bibfnamefont {S.}~\bibnamefont {Lloyd}},\ }\bibfield
   {title} {\bibinfo {title} {Quantum algorithm for linear systems of
  equations},\ }\href {https://doi.org/10.1103/PhysRevLett.103.150502}
  {\bibfield  {journal} {\bibinfo  {journal} {Phys. Rev. Lett.}\ }\textbf
  {\bibinfo {volume} {103}},\ \bibinfo {pages} {150502} (\bibinfo {year}
  {2009})}\BibitemShut {NoStop}%
\bibitem [{\citenamefont {{O'Brien}}\ \emph {et~al.}(2009)\citenamefont
  {{O'Brien}}, \citenamefont {Furusawa},\ and\ \citenamefont
  {Vu{\v{c}}kovi{\'c}}}]{obrienfurusawavuckovic:photonic}%
  \BibitemOpen
  \bibfield  {author} {\bibinfo {author} {\bibfnamefont {J.~L.}\ \bibnamefont
  {{O'Brien}}}, \bibinfo {author} {\bibfnamefont {A.}~\bibnamefont
  {Furusawa}},\ and\ \bibinfo {author} {\bibfnamefont {J.}~\bibnamefont
  {Vu{\v{c}}kovi{\'c}}},\ }\bibfield  {title} {\bibinfo {title} {Photonic
  quantum technologies},\ }\href {https://doi.org/10.1038/nphoton.2009.229}
  {\bibfield  {journal} {\bibinfo  {journal} {Nature Photonics}\ }\textbf
  {\bibinfo {volume} {3}},\ \bibinfo {pages} {687} (\bibinfo {year}
  {2009})}\BibitemShut {NoStop}%
\bibitem [{\citenamefont {Devoret}\ and\ \citenamefont
  {Schoelkopf}(2013)}]{devoretschoelkopf:superconducting}%
  \BibitemOpen
  \bibfield  {author} {\bibinfo {author} {\bibfnamefont {M.~H.}\ \bibnamefont
  {Devoret}}\ and\ \bibinfo {author} {\bibfnamefont {R.~J.}\ \bibnamefont
  {Schoelkopf}},\ }\bibfield  {title} {\bibinfo {title} {Superconducting
  circuits for quantum information: an outlook},\ }\href
  {https://doi.org/10.1126/science.1231930} {\bibfield  {journal} {\bibinfo
  {journal} {Science}\ }\textbf {\bibinfo {volume} {339}},\ \bibinfo {pages}
  {1169} (\bibinfo {year} {2013})}\BibitemShut {NoStop}%
\bibitem [{\citenamefont {Horodecki}\ \emph {et~al.}(2009)\citenamefont
  {Horodecki}, \citenamefont {Horodecki}, \citenamefont {Horodecki},\ and\
  \citenamefont {Horodecki}}]{horodecki:entanglement}%
  \BibitemOpen
  \bibfield  {author} {\bibinfo {author} {\bibfnamefont {R.}~\bibnamefont
  {Horodecki}}, \bibinfo {author} {\bibfnamefont {P.}~\bibnamefont
  {Horodecki}}, \bibinfo {author} {\bibfnamefont {M.}~\bibnamefont
  {Horodecki}},\ and\ \bibinfo {author} {\bibfnamefont {K.}~\bibnamefont
  {Horodecki}},\ }\bibfield  {title} {\bibinfo {title} {Quantum entanglement},\
  }\href {http://doi.org/10.1103/RevModPhys.81.865} {\bibfield  {journal}
  {\bibinfo  {journal} {Rev. Mod. Phys.}\ }\textbf {\bibinfo {volume} {81}},\
  \bibinfo {pages} {865} (\bibinfo {year} {2009})}\BibitemShut {NoStop}%
\bibitem [{\citenamefont {Raussendorf}\ and\ \citenamefont
  {Briegel}(2001)}]{raussendorfbriegel:mbqc}%
  \BibitemOpen
  \bibfield  {author} {\bibinfo {author} {\bibfnamefont {R.}~\bibnamefont
  {Raussendorf}}\ and\ \bibinfo {author} {\bibfnamefont {H.~J.}\ \bibnamefont
  {Briegel}},\ }\bibfield  {title} {\bibinfo {title} {A one-way quantum
  computer},\ }\href {https://doi.org/10.1103/PhysRevLett.86.5188} {\bibfield
  {journal} {\bibinfo  {journal} {Phys. Rev. Lett.}\ }\textbf {\bibinfo
  {volume} {86}},\ \bibinfo {pages} {5188} (\bibinfo {year}
  {2001})}\BibitemShut {NoStop}%
\bibitem [{\citenamefont {Deutsch}(1989)}]{deutsch:networks}%
  \BibitemOpen
  \bibfield  {author} {\bibinfo {author} {\bibfnamefont {D.}~\bibnamefont
  {Deutsch}},\ }\bibfield  {title} {\bibinfo {title} {Quantum computational
  networks},\ }\href {https://doi.org/10.1098/rspa.1989.0099} {\bibfield
  {journal} {\bibinfo  {journal} {Proc. Roy. Soc. A}\ }\textbf {\bibinfo
  {volume} {425}},\ \bibinfo {pages} {73} (\bibinfo {year} {1989})}\BibitemShut
  {NoStop}%
\bibitem [{\citenamefont {Regev}(2004)}]{USV}%
  \BibitemOpen
  \bibfield  {author} {\bibinfo {author} {\bibfnamefont {O.}~\bibnamefont
  {Regev}},\ }\bibfield  {title} {\bibinfo {title} {Quantum computation and
  lattice problems},\ }\href {https://doi.org/10.1137/S0097539703440678}
  {\bibfield  {journal} {\bibinfo  {journal} {SIAM J. Comput.}\ }\textbf
  {\bibinfo {volume} {33}},\ \bibinfo {pages} {738} (\bibinfo {year}
  {2004})}\BibitemShut {NoStop}%
\bibitem [{Note1()}]{Note1}%
  \BibitemOpen
  \bibinfo {note} {Extracted from Quipper USV-R implementation~\cite
  {quipperonline}.}\BibitemShut {Stop}%
\bibitem [{\citenamefont {Preskill}(2018)}]{NISQ}%
  \BibitemOpen
  \bibfield  {author} {\bibinfo {author} {\bibfnamefont {J.}~\bibnamefont
  {Preskill}},\ }\bibfield  {title} {\bibinfo {title} {Quantum {C}omputing in
  the {NISQ} era and beyond},\ }\href
  {https://doi.org/10.22331/q-2018-08-06-79} {\bibfield  {journal} {\bibinfo
  {journal} {{Quantum}}\ }\textbf {\bibinfo {volume} {2}},\ \bibinfo {pages}
  {79} (\bibinfo {year} {2018})}\BibitemShut {NoStop}%
\bibitem [{\citenamefont {{van Meter}}\ and\ \citenamefont
  {Devitt}(2016)}]{vanmeterdevitt:survey}%
  \BibitemOpen
  \bibfield  {author} {\bibinfo {author} {\bibfnamefont {R.}~\bibnamefont {{van
  Meter}}}\ and\ \bibinfo {author} {\bibfnamefont {S.~J.}\ \bibnamefont
  {Devitt}},\ }\bibfield  {title} {\bibinfo {title} {Local and distributed
  quantum computation},\ }\href {https://doi.org/10.1109/MC.2016.291}
  {\bibfield  {journal} {\bibinfo  {journal} {IEEE Computer}\ }\textbf
  {\bibinfo {volume} {49}},\ \bibinfo {pages} {31} (\bibinfo {year}
  {2016})}\BibitemShut {NoStop}%
\bibitem [{\citenamefont {Monroe}\ \emph {et~al.}(2014)\citenamefont {Monroe},
  \citenamefont {Raussendorf}, \citenamefont {Ruthven}, \citenamefont {Brown},
  \citenamefont {Maunz}, \citenamefont {Duan},\ and\ \citenamefont
  {Kim}}]{HierarchichalIonTrap}%
  \BibitemOpen
  \bibfield  {author} {\bibinfo {author} {\bibfnamefont {C.}~\bibnamefont
  {Monroe}}, \bibinfo {author} {\bibfnamefont {R.}~\bibnamefont {Raussendorf}},
  \bibinfo {author} {\bibfnamefont {A.}~\bibnamefont {Ruthven}}, \bibinfo
  {author} {\bibfnamefont {K.~R.}\ \bibnamefont {Brown}}, \bibinfo {author}
  {\bibfnamefont {P.}~\bibnamefont {Maunz}}, \bibinfo {author} {\bibfnamefont
  {L.-M.}\ \bibnamefont {Duan}},\ and\ \bibinfo {author} {\bibfnamefont
  {J.}~\bibnamefont {Kim}},\ }\bibfield  {title} {\bibinfo {title} {Large-scale
  modular quantum-computer architecture with atomic memory and photonic
  interconnects},\ }\href {https://doi.org/10.1103/PhysRevA.89.022317}
  {\bibfield  {journal} {\bibinfo  {journal} {Phys. Rev. A}\ }\textbf {\bibinfo
  {volume} {89}},\ \bibinfo {pages} {022317} (\bibinfo {year}
  {2014})}\BibitemShut {NoStop}%
\bibitem [{\citenamefont {van Meter}\ \emph {et~al.}(2010)\citenamefont {van
  Meter}, \citenamefont {Ladd}, \citenamefont {Fowler},\ and\ \citenamefont
  {Yamamoto}}]{HierarchichalQuantumDot}%
  \BibitemOpen
  \bibfield  {author} {\bibinfo {author} {\bibfnamefont {R.}~\bibnamefont {van
  Meter}}, \bibinfo {author} {\bibfnamefont {T.~D.}\ \bibnamefont {Ladd}},
  \bibinfo {author} {\bibfnamefont {A.~G.}\ \bibnamefont {Fowler}},\ and\
  \bibinfo {author} {\bibfnamefont {Y.}~\bibnamefont {Yamamoto}},\ }\bibfield
  {title} {\bibinfo {title} {Distributed quantum computation architecture using
  semiconductor nanophotonics},\ }\href
  {https://doi.org/10.1142/S0219749910006435} {\bibfield  {journal} {\bibinfo
  {journal} {Int.\ J.\ Q.\ Inf.}\ }\textbf {\bibinfo {volume} {8}},\ \bibinfo
  {pages} {295} (\bibinfo {year} {2010})}\BibitemShut {NoStop}%
\bibitem [{\citenamefont {Ballance}\ \emph {et~al.}(2015)\citenamefont
  {Ballance} \emph {et~al.}}]{nqit}%
  \BibitemOpen
  \bibfield  {author} {\bibinfo {author} {\bibfnamefont {C.~J.}\ \bibnamefont
  {Ballance}} \emph {et~al.},\ }\bibfield  {title} {\bibinfo {title} {Hybrid
  quantum logic and a test of bell’s inequality using two different atomic
  isotopes},\ }\href {https://doi.org/10.1038/nature16184} {\bibfield
  {journal} {\bibinfo  {journal} {Nature}\ }\textbf {\bibinfo {volume} {528}},\
  \bibinfo {pages} {384} (\bibinfo {year} {2015})}\BibitemShut {NoStop}%
\bibitem [{\citenamefont {Blok}\ \emph {et~al.}(2015)\citenamefont {Blok},
  \citenamefont {Kalb}, \citenamefont {Reiserer}, \citenamefont {Taminiau},\
  and\ \citenamefont {Hanson}}]{delft}%
  \BibitemOpen
  \bibfield  {author} {\bibinfo {author} {\bibfnamefont {M.~S.}\ \bibnamefont
  {Blok}}, \bibinfo {author} {\bibfnamefont {N.}~\bibnamefont {Kalb}}, \bibinfo
  {author} {\bibfnamefont {A.}~\bibnamefont {Reiserer}}, \bibinfo {author}
  {\bibfnamefont {T.~H.}\ \bibnamefont {Taminiau}},\ and\ \bibinfo {author}
  {\bibfnamefont {R.}~\bibnamefont {Hanson}},\ }\bibfield  {title} {\bibinfo
  {title} {Towards quantum networks of single spins: analysis of a quantum
  memory with an optical interface in diamond},\ }\href@noop {} {\bibfield
  {journal} {\bibinfo  {journal} {Faraday Discuss.}\ }\textbf {\bibinfo
  {volume} {184}},\ \bibinfo {pages} {173} (\bibinfo {year}
  {2015})}\BibitemShut {NoStop}%
\bibitem [{qua()}]{quantuminternetalliance}%
  \BibitemOpen
  \href@noop {} {}\bibinfo {note}
  {\href{http://quantum-internet.team}{quantum-internet.team}}\BibitemShut
  {NoStop}%
\bibitem [{\citenamefont {Wehner}\ \emph {et~al.}(2018)\citenamefont {Wehner},
  \citenamefont {Elkouss},\ and\ \citenamefont
  {Hanson}}]{wehnerelkousshanson:quantuminternet}%
  \BibitemOpen
  \bibfield  {author} {\bibinfo {author} {\bibfnamefont {S.}~\bibnamefont
  {Wehner}}, \bibinfo {author} {\bibfnamefont {D.}~\bibnamefont {Elkouss}},\
  and\ \bibinfo {author} {\bibfnamefont {R.}~\bibnamefont {Hanson}},\
  }\bibfield  {title} {\bibinfo {title} {Quantum internet: a vision for the
  road ahead},\ }\href {https://doi.org/10.1126/science.aam9288} {\bibfield
  {journal} {\bibinfo  {journal} {Science}\ }\textbf {\bibinfo {volume}
  {362}},\ \bibinfo {pages} {eaam9288} (\bibinfo {year} {2018})}\BibitemShut
  {NoStop}%
\bibitem [{\citenamefont {Schoute}\ \emph {et~al.}(2016)\citenamefont
  {Schoute}, \citenamefont {Mancinska}, \citenamefont {Islam}, \citenamefont
  {Kerenidis},\ and\ \citenamefont {Wehner}}]{schouteetal:routing}%
  \BibitemOpen
  \bibfield  {author} {\bibinfo {author} {\bibfnamefont {E.}~\bibnamefont
  {Schoute}}, \bibinfo {author} {\bibfnamefont {L.}~\bibnamefont {Mancinska}},
  \bibinfo {author} {\bibfnamefont {T.}~\bibnamefont {Islam}}, \bibinfo
  {author} {\bibfnamefont {I.}~\bibnamefont {Kerenidis}},\ and\ \bibinfo
  {author} {\bibfnamefont {S.}~\bibnamefont {Wehner}},\ }\bibfield  {title}
  {\bibinfo {title} {Shortcuts to quantum network routing},\ }\href
  {https://arxiv.org/abs/1610.05238} {\bibfield  {journal} {\bibinfo  {journal}
  {arXiv:1610.05238}\ } (\bibinfo {year} {2016})}\BibitemShut {NoStop}%
\bibitem [{\citenamefont {Hahn}\ \emph {et~al.}(2018)\citenamefont {Hahn},
  \citenamefont {Pappa},\ and\ \citenamefont
  {Eisert}}]{hahnpappeeisert:routing}%
  \BibitemOpen
  \bibfield  {author} {\bibinfo {author} {\bibfnamefont {F.}~\bibnamefont
  {Hahn}}, \bibinfo {author} {\bibfnamefont {A.}~\bibnamefont {Pappa}},\ and\
  \bibinfo {author} {\bibfnamefont {J.}~\bibnamefont {Eisert}},\ }\bibfield
  {title} {\bibinfo {title} {Quantum network routing and local
  complementation},\ }\href {https://arxiv.org/abs/1805.04559} {\bibfield
  {journal} {\bibinfo  {journal} {arXiv:1805.04559}\ } (\bibinfo {year}
  {2018})}\BibitemShut {NoStop}%
\bibitem [{\citenamefont {Akhremtsev}\ \emph {et~al.}(2017)\citenamefont
  {Akhremtsev}, \citenamefont {Heuer}, \citenamefont {Sanders},\ and\
  \citenamefont {Schlag}}]{KaHyPar}%
  \BibitemOpen
  \bibfield  {author} {\bibinfo {author} {\bibfnamefont {Y.}~\bibnamefont
  {Akhremtsev}}, \bibinfo {author} {\bibfnamefont {T.}~\bibnamefont {Heuer}},
  \bibinfo {author} {\bibfnamefont {P.}~\bibnamefont {Sanders}},\ and\ \bibinfo
  {author} {\bibfnamefont {S.}~\bibnamefont {Schlag}},\ }\bibfield  {title}
  {\bibinfo {title} {{Engineering a direct k-way Hypergraph Partitioning
  Algorithm}},\ }\href {https://doi.org/10.1137/1.9781611974768.3} {\bibfield
  {journal} {\bibinfo  {journal} {Proc.\ Alg.\ Eng.\ Exp.\ (ALENEX)}\ }\textbf
  {\bibinfo {volume} {19}},\ \bibinfo {pages} {28} (\bibinfo {year}
  {2017})}\BibitemShut {NoStop}%
\bibitem [{\citenamefont {{\c{C}}ataly{\"u}rek}\ and\ \citenamefont
  {Aykanat}(2011)}]{PaToH}%
  \BibitemOpen
  \bibfield  {author} {\bibinfo {author} {\bibfnamefont {{\"U}.}~\bibnamefont
  {{\c{C}}ataly{\"u}rek}}\ and\ \bibinfo {author} {\bibfnamefont
  {C.}~\bibnamefont {Aykanat}},\ }\bibinfo {title} {Patoh (partitioning tool
  for hypergraphs)},\ in\ \href {https://doi.org/10.1007/978-0-387-09766-4_93}
  {\emph {\bibinfo {booktitle} {Encyclopedia of Parallel Computing}}},\
  \bibinfo {editor} {edited by\ \bibinfo {editor} {\bibfnamefont
  {D.}~\bibnamefont {Padua}}}\ (\bibinfo  {publisher} {Springer US},\ \bibinfo
  {address} {Boston, MA},\ \bibinfo {year} {2011})\ pp.\ \bibinfo {pages}
  {1479--1487}\BibitemShut {NoStop}%
\bibitem [{\citenamefont {Bennett}\ \emph {et~al.}(1996)\citenamefont
  {Bennett}, \citenamefont {{DiVincenzo}}, \citenamefont {Smolin},\ and\
  \citenamefont {Wootters}}]{bennettdivincenzosmolinwooters:distillation}%
  \BibitemOpen
  \bibfield  {author} {\bibinfo {author} {\bibfnamefont {C.~H.}\ \bibnamefont
  {Bennett}}, \bibinfo {author} {\bibfnamefont {D.~P.}\ \bibnamefont
  {{DiVincenzo}}}, \bibinfo {author} {\bibfnamefont {J.~A.}\ \bibnamefont
  {Smolin}},\ and\ \bibinfo {author} {\bibfnamefont {W.~K.}\ \bibnamefont
  {Wootters}},\ }\bibfield  {title} {\bibinfo {title} {Mixed state entanglement
  and quantum error correction},\ }\href
  {https://doi.org/10.1103/PhysRevA.54.3824} {\bibfield  {journal} {\bibinfo
  {journal} {Phys. Rev. A}\ }\textbf {\bibinfo {volume} {54}},\ \bibinfo
  {pages} {3824} (\bibinfo {year} {1996})}\BibitemShut {NoStop}%
\bibitem [{\citenamefont {Cirac}\ \emph {et~al.}(1999)\citenamefont {Cirac},
  \citenamefont {Ekert}, \citenamefont {Huelga},\ and\ \citenamefont
  {Macchiavello}}]{NoisyChannels}%
  \BibitemOpen
  \bibfield  {author} {\bibinfo {author} {\bibfnamefont {J.~I.}\ \bibnamefont
  {Cirac}}, \bibinfo {author} {\bibfnamefont {A.~K.}\ \bibnamefont {Ekert}},
  \bibinfo {author} {\bibfnamefont {S.~F.}\ \bibnamefont {Huelga}},\ and\
  \bibinfo {author} {\bibfnamefont {C.}~\bibnamefont {Macchiavello}},\
  }\bibfield  {title} {\bibinfo {title} {Distributed quantum computation over
  noisy channels},\ }\href {https://doi.org/10.1103/PhysRevA.59.4249}
  {\bibfield  {journal} {\bibinfo  {journal} {Physical Review A}\ }\textbf
  {\bibinfo {volume} {59}},\ \bibinfo {pages} {4249} (\bibinfo {year}
  {1999})}\BibitemShut {NoStop}%
\bibitem [{\citenamefont {Simon}\ and\ \citenamefont
  {Irvine}(2003)}]{EbitGenHeralded}%
  \BibitemOpen
  \bibfield  {author} {\bibinfo {author} {\bibfnamefont {C.}~\bibnamefont
  {Simon}}\ and\ \bibinfo {author} {\bibfnamefont {W.~T.~M.}\ \bibnamefont
  {Irvine}},\ }\bibfield  {title} {\bibinfo {title} {Robust long-distance
  entanglement and a loophole-free bell test with ions and photons},\
  }\href@noop {} {\bibfield  {journal} {\bibinfo  {journal} {Physical review
  letters}\ }\textbf {\bibinfo {volume} {91}},\ \bibinfo {pages} {110405}
  (\bibinfo {year} {2003})}\BibitemShut {NoStop}%
\bibitem [{\citenamefont {Dawson}\ and\ \citenamefont
  {Nielsen}(2015)}]{SolovayKitaev}%
  \BibitemOpen
  \bibfield  {author} {\bibinfo {author} {\bibfnamefont {C.~M.}\ \bibnamefont
  {Dawson}}\ and\ \bibinfo {author} {\bibfnamefont {M.~A.}\ \bibnamefont
  {Nielsen}},\ }\bibfield  {title} {\bibinfo {title} {{The {S}olovay-{K}itaev
  algorithm}},\ }\href {https://www.arxiv.org/abs/quant-ph/0505030} {\bibfield
  {journal} {\bibinfo  {journal} {arXiv:quant-ph/0505030}\ } (\bibinfo {year}
  {2015})}\BibitemShut {NoStop}%
\bibitem [{\citenamefont {Yimsiriwattana}\ and\ \citenamefont {{Lomonaco
  Jr}}(2005)}]{Nonlocal}%
  \BibitemOpen
  \bibfield  {author} {\bibinfo {author} {\bibfnamefont {A.}~\bibnamefont
  {Yimsiriwattana}}\ and\ \bibinfo {author} {\bibfnamefont {S.~J.}\
  \bibnamefont {{Lomonaco Jr}}},\ }\bibfield  {title} {\bibinfo {title}
  {Generalized {GHZ} states and distributed quantum computing},\ }\href
  {https://arxiv.org/abs/quant-ph/0402148} {\bibfield  {journal} {\bibinfo
  {journal} {AMS Cont. Math.}\ }\textbf {\bibinfo {volume} {381}},\ \bibinfo
  {pages} {131} (\bibinfo {year} {2005})}\BibitemShut {NoStop}%
\bibitem [{\citenamefont {Gottesman}\ and\ \citenamefont
  {Chuang}(1999)}]{Gottesman}%
  \BibitemOpen
  \bibfield  {author} {\bibinfo {author} {\bibfnamefont {D.}~\bibnamefont
  {Gottesman}}\ and\ \bibinfo {author} {\bibfnamefont {I.~L.}\ \bibnamefont
  {Chuang}},\ }\bibfield  {title} {\bibinfo {title} {Demonstrating the
  viability of universal quantum computation using teleportation and
  single-qubit operations},\ }\href@noop {} {\bibfield  {journal} {\bibinfo
  {journal} {Nature}\ }\textbf {\bibinfo {volume} {402}},\ \bibinfo {pages}
  {390} (\bibinfo {year} {1999})}\BibitemShut {NoStop}%
\bibitem [{\citenamefont {Bennett}\ \emph {et~al.}(1993)\citenamefont
  {Bennett}, \citenamefont {Brassard}, \citenamefont {Cr{\'e}peau},
  \citenamefont {Jozsa}, \citenamefont {Peres},\ and\ \citenamefont
  {Wootters}}]{Teleportation}%
  \BibitemOpen
  \bibfield  {author} {\bibinfo {author} {\bibfnamefont {C.~H.}\ \bibnamefont
  {Bennett}}, \bibinfo {author} {\bibfnamefont {G.}~\bibnamefont {Brassard}},
  \bibinfo {author} {\bibfnamefont {C.}~\bibnamefont {Cr{\'e}peau}}, \bibinfo
  {author} {\bibfnamefont {R.}~\bibnamefont {Jozsa}}, \bibinfo {author}
  {\bibfnamefont {A.}~\bibnamefont {Peres}},\ and\ \bibinfo {author}
  {\bibfnamefont {W.~K.}\ \bibnamefont {Wootters}},\ }\bibfield  {title}
  {\bibinfo {title} {Teleporting an unknown quantum state via dual classical
  and einstein-podolsky-rosen channels},\ }\href@noop {} {\bibfield  {journal}
  {\bibinfo  {journal} {Physical review letters}\ }\textbf {\bibinfo {volume}
  {70}},\ \bibinfo {pages} {1895} (\bibinfo {year} {1993})}\BibitemShut
  {NoStop}%
\bibitem [{\citenamefont {Childs}\ \emph {et~al.}(2019)\citenamefont {Childs},
  \citenamefont {Schoute},\ and\ \citenamefont {Unsal}}]{ChildsSWAPs}%
  \BibitemOpen
  \bibfield  {author} {\bibinfo {author} {\bibfnamefont {A.~M.}\ \bibnamefont
  {Childs}}, \bibinfo {author} {\bibfnamefont {E.}~\bibnamefont {Schoute}},\
  and\ \bibinfo {author} {\bibfnamefont {C.~M.}\ \bibnamefont {Unsal}},\
  }\bibfield  {title} {\bibinfo {title} {Circuit transformations for quantum
  architectures},\ }\href@noop {} {\bibfield  {journal} {\bibinfo  {journal}
  {arXiv preprint arXiv:1902.09102}\ } (\bibinfo {year} {2019})}\BibitemShut
  {NoStop}%
\bibitem [{\citenamefont {Kissinger}\ and\ \citenamefont
  {de~Griend}(2019)}]{KissingerSteinerTrees}%
  \BibitemOpen
  \bibfield  {author} {\bibinfo {author} {\bibfnamefont {A.}~\bibnamefont
  {Kissinger}}\ and\ \bibinfo {author} {\bibfnamefont {A.~M.-v.}\ \bibnamefont
  {de~Griend}},\ }\bibfield  {title} {\bibinfo {title} {Cnot circuit extraction
  for topologically-constrained quantum memories},\ }\href@noop {} {\bibfield
  {journal} {\bibinfo  {journal} {arXiv preprint arXiv:1904.00633}\ } (\bibinfo
  {year} {2019})}\BibitemShut {NoStop}%
\bibitem [{\citenamefont {Nash}\ \emph {et~al.}(2019)\citenamefont {Nash},
  \citenamefont {Gheorghiu},\ and\ \citenamefont {Mosca}}]{NashSteinerTrees}%
  \BibitemOpen
  \bibfield  {author} {\bibinfo {author} {\bibfnamefont {B.}~\bibnamefont
  {Nash}}, \bibinfo {author} {\bibfnamefont {V.}~\bibnamefont {Gheorghiu}},\
  and\ \bibinfo {author} {\bibfnamefont {M.}~\bibnamefont {Mosca}},\ }\bibfield
   {title} {\bibinfo {title} {Quantum circuit optimizations for nisq
  architectures},\ }\href@noop {} {\bibfield  {journal} {\bibinfo  {journal}
  {arXiv preprint arXiv:1904.01972}\ } (\bibinfo {year} {2019})}\BibitemShut
  {NoStop}%
\bibitem [{\citenamefont {Zomorodi-Moghadam}\ \emph {et~al.}(2018)\citenamefont
  {Zomorodi-Moghadam}, \citenamefont {Houshmand},\ and\ \citenamefont
  {Houshmand}}]{PrevSol}%
  \BibitemOpen
  \bibfield  {author} {\bibinfo {author} {\bibfnamefont {M.}~\bibnamefont
  {Zomorodi-Moghadam}}, \bibinfo {author} {\bibfnamefont {M.}~\bibnamefont
  {Houshmand}},\ and\ \bibinfo {author} {\bibfnamefont {M.}~\bibnamefont
  {Houshmand}},\ }\bibfield  {title} {\bibinfo {title} {Optimizing
  teleportation cost in distributed quantum circuits},\ }\href
  {https://doi.org/10.1007/s10773-017-3618-x} {\bibfield  {journal} {\bibinfo
  {journal} {International Journal of Theoretical Physics}\ }\textbf {\bibinfo
  {volume} {57}},\ \bibinfo {pages} {848} (\bibinfo {year} {2018})}\BibitemShut
  {NoStop}%
\bibitem [{\citenamefont {Paler}\ \emph {et~al.}(2016)\citenamefont {Paler},
  \citenamefont {Wille},\ and\ \citenamefont {Devitt}}]{WireRecycling}%
  \BibitemOpen
  \bibfield  {author} {\bibinfo {author} {\bibfnamefont {A.}~\bibnamefont
  {Paler}}, \bibinfo {author} {\bibfnamefont {R.}~\bibnamefont {Wille}},\ and\
  \bibinfo {author} {\bibfnamefont {S.~J.}\ \bibnamefont {Devitt}},\ }\bibfield
   {title} {\bibinfo {title} {Wire recycling for quantum circuit
  optimization},\ }\href@noop {} {\bibfield  {journal} {\bibinfo  {journal}
  {Physical Review A}\ }\textbf {\bibinfo {volume} {94}},\ \bibinfo {pages}
  {042337} (\bibinfo {year} {2016})}\BibitemShut {NoStop}%
\bibitem [{\citenamefont {Markov}\ and\ \citenamefont
  {Shi}(2008)}]{ClassicalSimulation}%
  \BibitemOpen
  \bibfield  {author} {\bibinfo {author} {\bibfnamefont {I.~L.}\ \bibnamefont
  {Markov}}\ and\ \bibinfo {author} {\bibfnamefont {Y.}~\bibnamefont {Shi}},\
  }\bibfield  {title} {\bibinfo {title} {Simulating quantum computation by
  contracting tensor networks},\ }\href@noop {} {\bibfield  {journal} {\bibinfo
   {journal} {SIAM Journal on Computing}\ }\textbf {\bibinfo {volume} {38}},\
  \bibinfo {pages} {963} (\bibinfo {year} {2008})}\BibitemShut {NoStop}%
\bibitem [{qui()}]{quipperonline}%
  \BibitemOpen
  \href@noop {} {}\bibinfo {note}
  {\href{https://www.mathstat.dal.ca/~selinger/quipper/doc}{www.mathstat.dal.ca/$\sim$selinger/quipper/doc}}\BibitemShut
  {NoStop}%
\bibitem [{pab()}]{pablogithub}%
  \BibitemOpen
  \href@noop {} {}\bibinfo {note}
  {\href{https://github.com/PabloAndresMartinez/Distributed}{github.com/PabloAndresMartinez/Distributed}}\BibitemShut
  {NoStop}%
\bibitem [{\citenamefont {Ambainis}\ \emph {et~al.}(2007)\citenamefont
  {Ambainis}, \citenamefont {Childs}, \citenamefont {Reichardt}, \citenamefont
  {Spalek},\ and\ \citenamefont {Zhang}}]{BFWalk}%
  \BibitemOpen
  \bibfield  {author} {\bibinfo {author} {\bibfnamefont {A.}~\bibnamefont
  {Ambainis}}, \bibinfo {author} {\bibfnamefont {A.~M.}\ \bibnamefont
  {Childs}}, \bibinfo {author} {\bibfnamefont {B.~W.}\ \bibnamefont
  {Reichardt}}, \bibinfo {author} {\bibfnamefont {R.}~\bibnamefont {Spalek}},\
  and\ \bibinfo {author} {\bibfnamefont {S.}~\bibnamefont {Zhang}},\ }\bibfield
   {title} {\bibinfo {title} {Any and-or formula of size n can be evaluated in
  time \(n^{1/2 + o(1)}\) on a quantum computer},\ }\href
  {https://doi.org/10.1109/FOCS.2007.57} {\bibfield  {journal} {\bibinfo
  {journal} {Found.\ Comp.\ Sci. (FoCS)}\ }\textbf {\bibinfo {volume} {48}},\
  \bibinfo {pages} {363} (\bibinfo {year} {2007})}\BibitemShut {NoStop}%
\bibitem [{\citenamefont {Childs}\ \emph {et~al.}(2003)\citenamefont {Childs},
  \citenamefont {Cleve}, \citenamefont {Deotto}, \citenamefont {Farhi},
  \citenamefont {Gutmann},\ and\ \citenamefont {Spielman}}]{BWT}%
  \BibitemOpen
  \bibfield  {author} {\bibinfo {author} {\bibfnamefont {A.~M.}\ \bibnamefont
  {Childs}}, \bibinfo {author} {\bibfnamefont {R.}~\bibnamefont {Cleve}},
  \bibinfo {author} {\bibfnamefont {E.}~\bibnamefont {Deotto}}, \bibinfo
  {author} {\bibfnamefont {E.}~\bibnamefont {Farhi}}, \bibinfo {author}
  {\bibfnamefont {S.}~\bibnamefont {Gutmann}},\ and\ \bibinfo {author}
  {\bibfnamefont {D.~A.}\ \bibnamefont {Spielman}},\ }\bibfield  {title}
  {\bibinfo {title} {{Exponential algorithmic speedup by quantum walk}},\
  }\href {https://doi.org/10.1145/780542.780552} {\bibfield  {journal}
  {\bibinfo  {journal} {Proc.\ Symp.\ Th.\ Comp.\ (STOC)}\ }\textbf {\bibinfo
  {volume} {35}},\ \bibinfo {pages} {59} (\bibinfo {year} {2003})}\BibitemShut
  {NoStop}%
\bibitem [{\citenamefont {Whitfield}\ \emph {et~al.}(2011)\citenamefont
  {Whitfield}, \citenamefont {Biamonte},\ and\ \citenamefont
  {Aspuru-Guzik}}]{GSE}%
  \BibitemOpen
  \bibfield  {author} {\bibinfo {author} {\bibfnamefont {J.~D.}\ \bibnamefont
  {Whitfield}}, \bibinfo {author} {\bibfnamefont {J.}~\bibnamefont
  {Biamonte}},\ and\ \bibinfo {author} {\bibfnamefont {A.}~\bibnamefont
  {Aspuru-Guzik}},\ }\bibfield  {title} {\bibinfo {title} {Simulation of
  electronic structure hamiltonians using quantum computers},\ }\href
  {https://doi.org/10.1080/00268976.2011.552441} {\bibfield  {journal}
  {\bibinfo  {journal} {Molecular Physics}\ }\textbf {\bibinfo {volume}
  {109}},\ \bibinfo {pages} {735} (\bibinfo {year} {2011})}\BibitemShut
  {NoStop}%
\bibitem [{\citenamefont {Shende}\ and\ \citenamefont
  {Markov}(2009)}]{ToffoliDecomp}%
  \BibitemOpen
  \bibfield  {author} {\bibinfo {author} {\bibfnamefont {V.~V.}\ \bibnamefont
  {Shende}}\ and\ \bibinfo {author} {\bibfnamefont {I.~L.}\ \bibnamefont
  {Markov}},\ }\bibfield  {title} {\bibinfo {title} {On the cnot-cost of
  toffoli gates},\ }\href {http://dl.acm.org/citation.cfm?id=2011791.2011799}
  {\bibfield  {journal} {\bibinfo  {journal} {Quantum Info. Comput.}\ }\textbf
  {\bibinfo {volume} {9}},\ \bibinfo {pages} {461} (\bibinfo {year}
  {2009})}\BibitemShut {NoStop}%
\bibitem [{\citenamefont {Lyaudet}(2010)}]{NP-hard}%
  \BibitemOpen
  \bibfield  {author} {\bibinfo {author} {\bibfnamefont {L.}~\bibnamefont
  {Lyaudet}},\ }\bibfield  {title} {\bibinfo {title} {{NP-hard and linear
  variants of hypergraph partitioning}},\ }\href
  {https://doi.org/https://doi.org/10.1016/j.tcs.2009.08.035} {\bibfield
  {journal} {\bibinfo  {journal} {Theoretical Computer Science}\ }\textbf
  {\bibinfo {volume} {411}},\ \bibinfo {pages} {10 } (\bibinfo {year}
  {2010})}\BibitemShut {NoStop}%
\bibitem [{\citenamefont {Cowtan}\ \emph {et~al.}(2019)\citenamefont {Cowtan},
  \citenamefont {Dilkes}, \citenamefont {Duncan}, \citenamefont {Simmons},\
  and\ \citenamefont {Sivarajah}}]{CQCReduction}%
  \BibitemOpen
  \bibfield  {author} {\bibinfo {author} {\bibfnamefont {A.}~\bibnamefont
  {Cowtan}}, \bibinfo {author} {\bibfnamefont {S.}~\bibnamefont {Dilkes}},
  \bibinfo {author} {\bibfnamefont {R.}~\bibnamefont {Duncan}}, \bibinfo
  {author} {\bibfnamefont {W.}~\bibnamefont {Simmons}},\ and\ \bibinfo {author}
  {\bibfnamefont {S.}~\bibnamefont {Sivarajah}},\ }\bibfield  {title} {\bibinfo
  {title} {Phase gadget synthesis for shallow circuits},\ }\href@noop {}
  {\bibfield  {journal} {\bibinfo  {journal} {Proc.\ 16th International
  Conference on Quantum Physics and Logic (QPL, Orange, California, June
  2019)}\ } (\bibinfo {year} {2019})}\BibitemShut {NoStop}%
\bibitem [{\citenamefont {Amy}\ \emph {et~al.}(2014)\citenamefont {Amy},
  \citenamefont {Maslov},\ and\ \citenamefont {Mosca}}]{TReduction}%
  \BibitemOpen
  \bibfield  {author} {\bibinfo {author} {\bibfnamefont {M.}~\bibnamefont
  {Amy}}, \bibinfo {author} {\bibfnamefont {D.}~\bibnamefont {Maslov}},\ and\
  \bibinfo {author} {\bibfnamefont {M.}~\bibnamefont {Mosca}},\ }\bibfield
  {title} {\bibinfo {title} {Polynomial-time t-depth optimization of clifford+
  t circuits via matroid partitioning},\ }\href@noop {} {\bibfield  {journal}
  {\bibinfo  {journal} {IEEE Transactions on Computer-Aided Design of
  Integrated Circuits and Systems}\ }\textbf {\bibinfo {volume} {33}},\
  \bibinfo {pages} {1476} (\bibinfo {year} {2014})}\BibitemShut {NoStop}%
\end{thebibliography}%

\section*{Appendix A}

In this appendix we prove the theorem presented in section~\ref{sec:reduction} and related results also discussed in that section.

\vspace{1ex}
\textbf{Theorem} \emph{Given a circuit, each of its possible distributed implementations (without altering the gateset or the gate order) corresponds to a unique partition of its hypergraph (given by Figure~\ref{code:buildHyp}) whose number of cuts is equivalent to the number of ebits required.}
\vspace{1ex}

\emph{Proof.} First, we provide a bijection between the trivial configurations:
\begin{itemize}
  \item a partition of the hypergraph where all vertices are in the same block corresponds one-to-one to 
  \item the whole circuit being executed in a single QPU.
\end{itemize}

Then, we extend the bijection to any partition/distribution by defining two primitive transformations for each problems, which allow us to move vertices around. The wire-primitive moves wire-vertices:
\begin{itemize}
  \item given a partition of the hypergraph, moving wire-vertex \(x\) to block \(i\) corresponds one-to-one to
  \item picking wire \(x\) and allocating it to QPU \(i\).
\end{itemize}
The \emph{CZ-primitive} moves CZ-vertices:
\begin{itemize}
  \item given a partition of the hypergraph, moving CZ-vertex \(\alpha\) to block \(i\) corresponds one-to-one to
  \item picking CZ gate \(\alpha\) and allocating it to be carried out in QPU \(i\).
\end{itemize}

Any partition/distribution can be described as a sequence of primitives: starting from the trivial configuration, first move all CZs to their corresponding block/QPU using the CZ-primitive once per CZ gate, then do the same for the wires using the wire-primitive. The one-to-one correspondence between primitives then gives us a bijection between the set of all possible distributions of the circuit and all possible partitions of its hypergraph. It remains to prove this bijection satisfies that the ebit count \(\lambda_e\) of a distribution and the cut count \(\lambda_c\) of its corresponding hypergraph partition are always equivalent \(\lambda_c = \lambda_e\).

\begin{enumerate}
  \item The trivial configuration of both problems has \(\lambda_c = 0 = \lambda_e\). We impose that the block/QPU where all vertices are allocated on the trivial configuration is an auxiliary one that will not hold any vertices/wires on the final partition/distribution. Thus, it is just an artifact to simplify the proof.
  \item By construction, each CZ-vertex is connected to exactly two hyperedges. When a CZ-primitive is applied, the number of cuts \(\lambda_c\) will increase by one if and only if, in the block where it is reallocated, there is no other CZ-vertex with whom it shares a hyperedge. The same happens for the ebit count \(\lambda_e\): if, in the QPU where it is reallocated, there is no CZ with whom it shares a wire, then an ebit is required to remotely access it, otherwise the channel already exists and no additional ebit is required. Thus, we may reallocate all CZ gates while preserving \(\lambda_c = \lambda_e\).
  \item Applying wire-primitives to the current configuration will always decrease \(\lambda_c\) and \(\lambda_e\). When wire-vertex \(x\) is reallocated to block \(i\), the number of cuts \(\lambda_c\) will decrease by one per hyperedge \(x\) shares with a CZ-vertex in block \(i\). The ebit count \(\lambda_e\) will decrease under the same circumstances, because the CZs corresponding to those CZ-vertices will be able to access the wire locally, and therefore will not require ebits to do so. Thus, we may reallocate all wires while preserving \(\lambda_c = \lambda_e\).
\end{enumerate} 

The strategy of allocating all CZ gates first and then allocating wires is chosen for simplicity. Although tedious, it is straightforward to check case by case that the argument above holds independently of the order at which we allocate CZ gates and wires. \begin{flushright}$\square$\end{flushright}

The following lemma shows that, in a similar way as to how the problem of quantum circuit distribution can be reduced to hypergraph partitioning, the dual notion is also true: hypergraph partitioning can be solved using a quantum circuit distributer. This insight has no direct application in practice, but it is valuable from a theoretical point of view, as stated in the corollary that follows.

\vspace{1ex}
\textbf{Lemma} \emph{The problem of hypergraph partitioning can be reduced to the problem of quantum circuit distribution.}
\vspace{1ex}

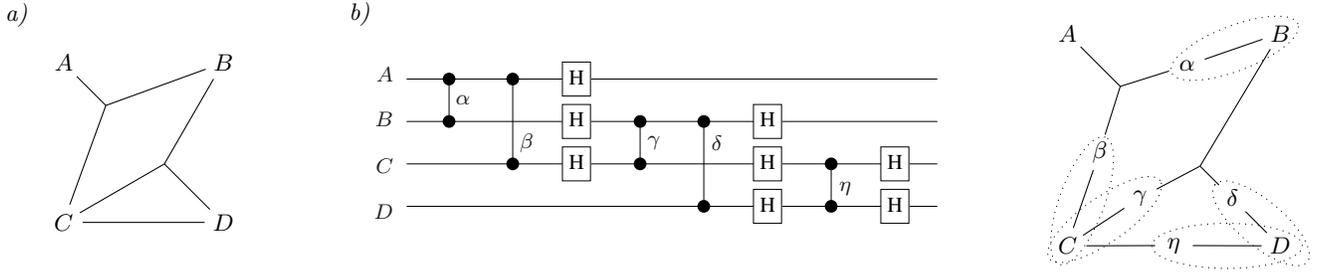
\begin{figure*}
\begin{tikzpicture}
  \node (input) {
    \begin{tikzpicture}
      \coordinate (auxC) at (135:7mm);
      \coordinate (auxB) at (315:4mm);
      \coordinate (A) at (135:15mm);
      \coordinate (B) at (45:15mm);
      \coordinate (C) at (225:15mm);
      \coordinate (D) at (315:15mm);
      \draw (auxC) -- (A);
      \draw (auxC) -- (B);
      \draw (auxC) -- (C);
      \draw (auxB) -- (B);
      \draw (auxB) -- (C);
      \draw (auxB) -- (D);
      \draw (C) -- (D);
      \node[circle, right=-2.5mm of A, fill=white, inner sep=0pt, minimum size=5mm] {\(A\)};
      \node[circle, right=-2.5mm of B, fill=white, inner sep=0pt, minimum size=5mm] {\(B\)};
      \node[circle, right=-2.5mm of C, fill=white, inner sep=0pt, minimum size=5mm] {\(C\)};
      \node[circle, right=-2.5mm of D, fill=white, inner sep=0pt, minimum size=5mm] {\(D\)};
    \end{tikzpicture}
  };
  \node[above left=0mm of input, font=\itshape] (labela) {a)};
  \node[right=10mm of input] (dummyCirc) {
    \begin{tikzpicture}
      \node (circuit) {\includegraphics[scale=1.6]{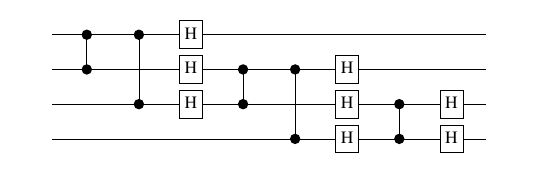}};
      \node[above left=7mm and -9mm of circuit.west, opacity=0.9] {\footnotesize \(A\)};
      \node[above left=1mm and -9mm of circuit.west, opacity=0.9] {\footnotesize \(B\)};
      \node[below left=1mm and -9mm of circuit.west, opacity=0.9] {\footnotesize \(C\)};
      \node[below left=7mm and -9mm of circuit.west, opacity=0.9] {\footnotesize \(D\)};
      \node[above right=4mm and 15mm of circuit.west, opacity=0.9] {\footnotesize \(\alpha\)};
      \node[right=23.5mm of circuit.west, opacity=0.9] {\footnotesize \(\beta\)};
      \node[right=40.5mm of circuit.west, opacity=0.9] {\footnotesize \(\gamma\)};
      \node[right=49mm of circuit.west, opacity=0.9] {\footnotesize \(\delta\)};
      \node[below right=4mm and 66mm of circuit.west, opacity=0.9] {\footnotesize \(\eta\)};
    \end{tikzpicture}
  };
  \node[above left=-2mm and -7mm of dummyCirc, font=\itshape] (labelb) {b)};
  \node[right=2mm of dummyCirc] (extended) {
    \begin{tikzpicture}      
      \coordinate (auxC) at (135:10mm);
      \coordinate (auxB) at (315:5mm);
      \coordinate (A) at (135:20mm);
      \coordinate (B) at (45:20mm);
      \coordinate (C) at (225:20mm);
      \coordinate (D) at (315:20mm);
      \coordinate (a) at (80:10mm);
      \coordinate (b) at (190:10mm);
      \coordinate (e) at (270:14.1mm);
      \coordinate (c) at (240:9mm);
      \coordinate (d) at (315:11mm);
      \coordinate[below left=2mm and 6mm of B] (Ba);
      \coordinate[above right=6mm and 2mm of C] (Cb);
      \coordinate[above right=3.5mm and 5mm of C] (Cc);
      \coordinate[above left=3mm and 2.5mm of D] (Dd);
      \coordinate[left=7mm of D] (De);
      \draw (auxC) -- (a);
      \draw (auxC) -- (b);
      \draw (auxC) -- (A);
      \draw (auxB) -- (B);
      \draw (auxB) -- (c);
      \draw (auxB) -- (d);
      \draw (C) -- (c);
      \draw (D) -- (d);
      \draw (B) -- (a);
      \draw (C) -- (b);
      \draw (D) -- (e);
      \draw (C) -- (e);
      \node[circle, right=-2.5mm of A, fill=white, inner sep=0pt, minimum size=5mm] {\(A\)};
      \node[circle, right=-2.5mm of B, fill=white, inner sep=0pt, minimum size=5mm] {\(B\)};
      \node[circle, right=-2.5mm of C, fill=white, inner sep=0pt, minimum size=5mm] {\(C\)};
      \node[circle, right=-2.5mm of D, fill=white, inner sep=0pt, minimum size=5mm] {\(D\)};
      \node[circle, right=-2.5mm of a, fill=white, inner sep=0pt, minimum size=5mm] {\(\alpha\)};
      \node[circle, right=-2.5mm of b, fill=white, inner sep=0pt, minimum size=5mm] {\(\beta\)};
      \node[circle, right=-2.5mm of c, fill=white, inner sep=0pt, minimum size=5mm] {\(\gamma\)};
      \node[circle, right=-2.5mm of d, fill=white, inner sep=0pt, minimum size=5mm] {\(\delta\)};
      \node[circle, right=-2.5mm of e, fill=white, inner sep=0pt, minimum size=5mm] {\(\eta\)};
      \draw[dotted, rotate around=(21:(Ba))] (Ba) ellipse (9mm and 3mm);
      \draw[dotted, rotate around=(70:(Cb))] (Cb) ellipse (9mm and 3mm);
      \draw[dotted, rotate around=(35:(Cc))] (Cc) ellipse (9mm and 3mm);
      \draw[dotted, rotate around=(140:(Dd))] (Dd) ellipse (8mm and 3mm);
      \draw[dotted] (De) ellipse (10mm and 3mm);    \end{tikzpicture}
  };
  \node[above left=0mm of extended, font=\itshape] (labelc) {c)};
\end{tikzpicture}
\caption{\emph{a)} An arbitrary input hypergraph. \emph{b)} A (not unique) dummy circuit built from the hypergraph. \emph{c)} The hypergraph obtained by applying the algorithm from Figure~\ref{code:buildHyp} on the dummy circuit, the hypergraph obtained is an extended version of the input one. We can retrieve the input hypergraph by merging vertices as indicated by the dotted ellipses.}
\label{fig:dualProcess}
\end{figure*}

\emph{Proof.} We need to show how an optimal partition of any hypergraph can be obtained by finding an optimal distribution of a dummy circuit. Given any hypergraph, create a circuit that has one wire per vertex in the hypergraph and, for each hyperedge \(h\): 
\begin{enumerate} 
  \item take the subset of vertices it reaches and the corresponding subset of wires \(\mathcal{W}_h\);
  \item pick (at random) one of these wires and apply CZ gates between it and each of the other wires in \(\mathcal{W}_h\);
  \item apply a Hadamard gate on each of the wires in \(\mathcal{W}_h\).
\end{enumerate}
Notice that this process takes a polynomial number of steps with respect to the number of vertices and hyperedges in the hypergraph. An example of a hypergraph and its resulting dummy circuit is given in Figure~\ref{fig:dualProcess}.

We then use the algorithm from Figure~\ref{code:buildHyp} to obtain a new hypergraph (see Figure~\ref{fig:dualProcess}c) which is similar to the original one, but not the same. We call this hypergraph the \emph{extended} hypergraph. Notice that the only difference is the addition of CZ-vertices and standard edges. If we merge each CZ-vertex with the wire-vertex the standard edge connects it to, the resulting hypergraph is the same as the one given as input (see Figure~\ref{fig:dualProcess}c). It is trivial to check that this will always be the case due to the way we build the dummy circuit. 

If an optimal distribution of the dummy circuit is found, we can use the bijection provided in the proof of the theorem above to obtain an optimal partition of the extended hypergraph. We can then remove the CZ-vertices by merging them, obtaining a partition of the input hypergraph. When the vertices to be merged are in the same block it is clear that merging does not affect the optimality of the partition. When they live in different blocks, reallocating the CZ-vertices so they are on the same block can never increase the cut count: either that cut is simply moved from the standard edge to the hyperedge, or the cut is no longer needed because another vertex in the hyperedge is already on that block. But because the partition of the extended hypergraph is optimal, this particular reallocation can not decrease the cut count either. Thus we may ignore the CZ-vertices altogether. The allocation of the rest of the vertices provides an optimal partition of the input hypergraph. \hfill$\square$

\vspace{1ex}
\textbf{Corollary} \emph{The problem of quantum circuit distribution is NP-hard.}
\vspace{1ex}

\emph{Proof.} The previous lemma shows that hypergraph partitioning can be reduced to this problem, with all required transformations having polynomial time complexity. As hypergraph partitioning is NP-hard~\cite{NP-hard}, it immediately follows by the definition of NP-hardness that the problem of quantum circuit distribution, as defined in this document, is NP-hard too. \hfill$\square$

\section*{Appendix B}

This appendix presents the procedure labelled \emph{pre-processing 3} in Section~\ref{sec:implementation}, that informs how the input circuit should be split into segments before distributing. The goal is that, whenever the qubit connectivity within the circuit changes dramatically, the circuit is divided into two segments: one ending at some point previous to that change and the other starting from that point onwards. The different segments are then distributed using the approach described in Section~\ref{sec:reduction}.

The procedure requires two user-defined parameters \(\omega \in \mathbb{N}\) and \(\Delta \in [0,1]\). First, the circuit is explored from left to right, splitting it into preliminary segments containing \(\omega\) many CZ gates each. Then, for each segment:
\begin{enumerate}
  \item obtain the hypergraph partitions of the current segment and the next one;
  \item obtain their discrepancy score \(\delta\) computed as in \eqref{eq:discrepancy}; 
  \item if the discrepancy $\delta$ is below the threshold $\Delta$, view both segments as a single one (i.e.\ merge them) and return to step 1; otherwise obtain the distributed circuit of the current segment and continue the procedure until all segments have been distributed.
\end{enumerate}
Once this process finishes, the distributed circuits are executed in the target DQA one after the other. This may require teleporting qubits~\cite{Teleportation} between QPUs when progressing from one segment to the next. Each qubit teleportation makes use of a single ebit; this cost has been taken into account in the figures and discussion of Section~\ref{sec:results}. 
The procedure may be modified so that parameter \(\Delta\) is not required: apply step 3 only when \(\delta\) is minimal and repeat the procedure until merging segments no longer decreases the ebit count. This modified version is the one used to obtain the results graphed in Section~\ref{sec:results}.

The the discrepancy score \(\delta \in [0,1]\) between two segments \(s\) and \(r\) is calculated as:
\begin{equation}
\delta = \sum_{w \in \mathcal{W}} \tau(w) \ \frac{\text{min}\{h_s(w), h_r(w)\}}{\text{min}\{H_s, H_r\}}
\label{eq:discrepancy}
\end{equation}
where \(\mathcal{W}\) is the set of all wires in the circuit, \(\tau(w)\) returns \(0\) if the wire is allocated to the same QPU in both segments and \(1\) otherwise, \(H_s\) returns the total number of hyperedges in the hypergraph of segment \(s\) (similarly for segment \(r\)) and \(h_s(w)\) returns the number of hyperedges that reach the vertex corresponding to wire \(w\) within the hypergraph of segment \(s\) (similarly for segment \(r\)). 

Different discrepancy scores could be used without changing any other aspect of the algorithm. Equation \eqref{eq:discrepancy} was the score that performed best among the different options we attempted. It moreover has an intuitive interpretation:
\begin{itemize}
  \item if a wire is allocated to the same QPU in both segments, that wire's contribution to discrepancy is null, which is why we multiply by \(\tau(w)\);
  \item \(h_s(w)\) estimates the wire's relevance in the segment connectivity and hence should be proportional to the discrepancy score;
  \item if a wire is allocated to different QPUs in each segment, but is almost never used in one of them (i.e.\ \(h_r(w) \approx 0\)), it would be relatively cheap to reallocate that wire to match the other segment, justifying the use of \((\text{min}\{h_s(w),h_r(w)\})\);
  \item to compare scores fairly they need to be normalized, which is why we divide by \((\text{min}\{H_s, H_r\})\).
\end{itemize}

This procedure uses the hypergraph partitioner multiple times. At first glance, that may seem to come at a great cost, as hypergraph partitioning is the most resource intensive routine in our approach. However, by splitting the circuit into segments, the hypergraph that is partitioned each time is much smaller than the hypergraph of the overall circuit. Considering that hypergraph partitioning is an NP-hard problem~\cite{NP-hard}, the reduction of the input size improves performance dramatically. In practice, when using KaHyPar~\cite{KaHyPar} (our choice of third-party hypergraph partitioner), we found out this performance improvement overcame the cost of running the partitioner multiple times.

\end{document}